\documentclass[twocolumn,preprintnumbers,amsmath,amssymb,prb]{revtex4}
\usepackage{graphicx}
\usepackage{amsmath}
\usepackage{amssymb}
\usepackage{appendix}
\graphicspath{{figures/}}

\begin{document}

\title{A linear combination of atomic orbitals (LCAO) model for deterministically placed acceptor arrays in silicon}
\author{Jianhua Zhu}\email{ucapjhz@ucl.ac.uk}
\author{Wei Wu}\email{wei.wu@ucl.ac.uk}
\author{Andrew J. Fisher}\email{andrew.fisher@ucl.ac.uk}
\affiliation{UCL Department of Physics and Astronomy and London Centre for Nanotechnology,\\
University College London, Gower Street, London WC1E 6BT, United Kingdom}

\begin{abstract}
We develop a tight-binding model based on linear combination of atomic orbitals (LCAO) methods to describe the electronic structure of arrays of acceptors, where the underlying basis states are derived from an effective-mass-theory solution for a single acceptor in either the spherical approximation or the cubic model. Our model allows for arbitrarily strong spin-orbit coupling in the valence band of the semiconductor.  We have studied pairs and dimerised linear chains of acceptors in silicon in the `independent-hole' approximation, and investigated the conditions for the existence of topological edge states in the chains. For the finite chain we find a complex interplay between electrostatic effects and the dimerisation, with the long-range Coulomb attraction of the hole to the acceptors splitting off states localised at the end acceptors from the rest of the chain.  A further pair of states then splits off from each band, to form a pair localised on the next-to-end acceptors, for one sense of the bond alternation and merges into the bulk bands for the other sense of the alternation. We confirm the topologically non-trivial nature of these next-to-end localised states by calculating the Zak phase.  We argue that for the more physically accessible case of one hole per acceptor these long-range electrostatic effects will be screened out; we show this by treating a simple phenomenologically screened model in which electrostatic contributions from beyond the nearest neighbours of acceptor each pair are removed.  Topological states are now found on the end acceptors of the chains.  In some cases the termination of the chain required to produce topological states is not the one expected on the basis of simple geometry (short versus long bonds); we argue this is because of a non-monotonic relationship between the bond length and the effective Hamiltonian matrix elements between the acceptors. 
\end{abstract}

\maketitle
\section{Introduction}\label{introduction}

Recently defects in semiconductors have aroused increasing interest owing to their applications in quantum simulation, quantum computation, and Terahertz radiation \cite{Sakai2005TO}. Donors are well understood and the exchange interaction of a pair of donors was studied within the Heitler-London approximation several decades ago \cite{Koiller2002EiSBQCA,Pantelides1974ToLSiSINRUaOM}. However, donors in indirect-gap systems suffer from the disadvantage that the oscillation of spin-spin interactions is large and not controllable, owing to the unavoidable interference between the conduction-band valleys of the host. Recently acceptors in tetrahedral semiconductors have attracted renewed attention, because of the absence of such multi-band interference in the valence band; this will lead to monotonic exchange and hopping interactions that are easier to control. However, owing to the $p$-orbital characters of the valence band, the spin-orbit interactions need to be taken into account from the outset.


Previously the electronic structures of a single acceptor in common semiconductors have been studied both theoretically and experimentally. Baldereschi and Lipari introduced the so-called `spherical model' \cite{Luttinger1956QToCRiSGT,Baldereschi1973SMoSASiS}, based on the effective-mass theory and including the cubic contributions either through perturbation theory \cite{Baldereschi1974CCttSMoSAS} or in an exact form \cite{Lipari1978IoASiS}. These calculations gave reasonably accurate predictions of the acceptor ionisation and excitation energies. Recently, Durst, et. al., computed the electronic structure and exchange interaction between two acceptors by adopting the spherical model and the Heitler-London approximation \cite{C.Durst2017HLMfAAIiDS}. They also investigated the interaction between acceptor pairs in the extreme long-range limit, where hopping of electrons is no longer relevant,  again using the spherical model \cite{C.Durst2019}; they argue that in this limit the interactions are dominated by electric quadrupole moments. On the other hand, experimental measurements of the optical transitions and spectra of acceptors in silicon have been performed \cite{Clauws1989OSoSISiGaS}; the coherence time of the excited state of acceptors in silicon has also been measured, showing promise for optically controlled $p-n$ devices \cite{Vinh2013TRDoSATiS}. The transport properties of boron in silicon, such as the conductivity and magnetoresistance have also been studied previously \cite{Dai1992ECoMSntMIT}. Recently, the readout and control of the spin-orbit state of two coupled acceptors in silicon was demonstrated experimentally, opening up another route to quantum computing and quantum information in silicon \cite{Litvinenko2015CCaDoOWiSwEaORO, Heijden2018RaCotSOSoTCAAiaST,Corna2018EDESRMbSVOCiaSQD,Crippa2018ESDbMMiSOQ,Crippa2019GRDRaCCoaSQiS}. Acceptor pairs in silicon have also been used for simulations of fermionic strongly-correlated many-body systems \cite{Salfi2016QSotHMwDAiS}, and this is particularly interesting in the context of the emerging field of deterministic doping \cite{Schofield2003APPoSDiS}. Although the surface chemistry needed for deterministic implantation of more complex structures has not yet been developed, it is timely to investigate the potential structures that could be produced, and the potential role of the spin-orbit interaction in their electronic properties.

Here we report a systematic study of ordered acceptor arrays in tetrahedral semiconductors by using a combination of a linear combination of atomic orbitals (LCAO) model and quantum chemistry calculations. We first computed the electronic structure of the single acceptor in silicon by using spherical and cubic models; our results confirm the significant improvement due to the inclusion of the cubic term, thus benchmarking the previous work. Based on these single-acceptor calculations, we have selected an appropriate basis set of single-acceptor electronic states and performed calculations on acceptor pairs and linear chains by using a linear combination of atomic orbitals (LCAO) approach within an independent-hole model. Our results suggest an interesting interplay between electrostatic effects and topological edge state in finite chains. The following discussion falls into three sections. In \S\ref{theory}, we will introduce the spherical model and the cubic model, and use them to develop our LCAO model for acceptor arrays. In \S\ref{result}, we will show our calculation results.  At the end, we will draw some general conclusions.

\section{Computational details}\label{theory}

\subsection{Single-acceptor problem}\label{single}

\subsubsection{Spherical model}\label{spherical}

Within effective mass theory \cite{Luttinger1956QToCRiSGT}, the Hamiltonian for an isolated acceptor contains spherical and non-spherical (cubic) parts. In many cases (although not in silicon), we can neglect the cubic terms and obtain the so-called spherical model \cite{Baldereschi1973SMoSASiS},  which can be significantly simplified by using rotational symmetry. In this paper, we take the general form of the spherical model (valid for arbitrary spin-orbit coupling) as follows,
\begin{equation}
\hat{H}_s=\frac{p^2}{\hbar^2}-\frac{2}{r}-\frac{\mu}{3\hbar^2}(P^{(2)}\bullet I^{(2)})+\frac{2}{3}\Delta(\frac{1}{2}-\vec{I}\bullet \vec{S})\label{e-2-1-1-1}
\end{equation}
where $p$ is the hole linear momentum operator, $\mu$ is the strength of the spherically symmetric heavy-hole light-hole coupling, and $\Delta$ is the spin-orbit coupling.  The tensor operators $P^{(2)}$ and $I^{(2)}$ are as defined in previous studies of acceptors \cite{Baldereschi1973SMoSASiS}: $P_{ik}=3p_ip_k-\delta_{ik}p^2$ contains the hole (linear) momenta, while $I_{ik}=\frac32(I_iI_k+I_kI_l)-\delta_{ik}I^2$ is built from the angular momentum operators $I_i$ of a spin-1 object (corresponding to the intrinsic orbital angular momentum of the $p$-orbitals comprising the valence band). $\vec{I}$ is the corresponding vector of spin-1 angular momentum operators, while $\vec{S}$ is the vector of spin-$\frac{1}{2}$ spin operators of the hole. In this model, we use the effective Rydberg $R_0=\frac{e^4m_0}{2\hbar^2\epsilon^2_0\gamma_1}$ and the effective Bohr radius $a_0=\frac{\hbar^2\epsilon^2_0\gamma_1}{e^4m_0}$ as units of energy and length, respectively \cite{Baldereschi1973SMoSASiS}, where $\epsilon_0$ and $m_0$ are the crystal dielectric constant and the free electron mass, respectively, and $\gamma_1$ is the parameter proposed by Luttinger for the description of the hole dispersion relation near the center of the Brillouin zone \cite{Luttinger1956QToCRiSGT}. For silicon, where the effective Rydberg $R_0=24.8\,\mathrm{meV}$ and $a_0=2.55\,\mathrm{nm}$, we have the strength of the spherical term $\mu=0.483$, and the valence band spin-orbit splitting $\Delta=1.774R_0$. We note that  the model is set up to describe electrons in the valence band, so the ground state for holes will appear at the top of the spectrum (i.e., with the largest positive eigenvalue). For convenience, we still use the common description to describe the energy states, for example, the the lowest state is the ground state.

The eigenstates of Equation (\ref{e-2-1-1-1}) have well defined values of the total angular momentum $\vec{F}=\vec{L}+\vec{I}+\vec{S}=\vec{L}+\vec{J}$, where $\vec{L}$ is the orbital angular momentum of the envelope function and $\vec{J}=\vec{I}+\vec{S}$ is the total intrinsic angular momentum of a valence-band electron.  Hence, they are characterised by quantum numbers $F$ and $m_F$. The spherical states used in the calculation are shown in Appendix \ref{states}. The heavy-hole light-hole mixing couples terms with $\Delta L=0,\pm 2$; its matrix elements can be obtained using the result below,
\begin{align}
\left\langle L',J',F,m_F\right|P^{(2)}\bullet I^{(2)}\left|L,J,F,m_F\right\rangle\notag\\
=(-1)^{L+J'+F}
\begin{Bmatrix}
F&J'&L'\\2&L&J
\end{Bmatrix}
\left(L'\right\|P^{(2)}\left\|L\right)\left(J'\right\|I^{(2)}\left\|J\right)\label{e-2-1-1-3}
\end{align}
where the term with $\left\{\right\}$ is the 6-j symbol, and $\left(J'\right\|I^{(2)}\left\|J\right)$ can be obtained by the formula
\begin{align}
\left(J'\right\|I^{(2)}\left\|J\right)=(-1)^{J+\frac{7}{2}}\sqrt{(2J+1)(2J'+1)}
\begin{Bmatrix}
1&J'&\frac{1}{2}\\J&1&2
\end{Bmatrix}\notag\\
\bullet\left(I'\right\|I^{(2)}\left\|I\right)\label{e-2-1-1-4}
\end{align}
Using these results, we find the differential equations satisfied by the radial parts of the wave functions. We then obtain the single-acceptor eigenstates and eigenenergies, which can be used as a basis for the further calculation of acceptor arrays in section \S\ref{pair} and \S\ref{chain}. 

\subsubsection{Cubic model}\label{cubic}

The cubic model takes the form
\begin{align}
\hat{H}_c=\hat{H}_s+\frac{\delta}{3\hbar^2}([P^{(2)}\times I^{(2)}]^{(4)}_4+\frac{\sqrt{70}}{5}[P^{(2)}\times I^{(2)}]^{(4)}_0\notag\\
+[P^{(2)}\times I^{(2)}]^{(4)}_{-4})\label{e-2-1-2-1}
\end{align}
where $\delta$ is the strength of the cubic term \cite{Baldereschi1974CCttSMoSAS} ($\delta=0.249$ for Si). Here we still use the effective Rydberg $R_0$ and the effective Bohr radius $a_0$ as units of energy and length, respectively \cite{Baldereschi1973SMoSASiS}, and $[A\times B]^{(l)}_m$ denotes component $m$ of the part of the spherical tensor product $A\times B$ having rank $l$.

The cubic term couples states with $\Delta m_F=0,\pm4$ \cite{Baldereschi1974CCttSMoSAS}, so the eigenstates are now labelled by irreducible representations of the cubic double group rather than by values of $F$.  There are 6 fermionic representations, $\Gamma^\pm_6,\Gamma^\pm_7,\Gamma^\pm_8$; states with these symmetries can be obtained by taking suitable linear combinations of states with spherical symmetry. We use an underlying basis of Gaussian orbitals of spherical symmetry up to a maximum of $L=3$ and $F=9/2$ (which we label as $F_\frac{9}{2}$ states, corresponding to the usual notation in atomic physics). We compute the matrix elements of the cubic terms in this basis of spherically symmetric states, using
\begin{align}
\left\langle L',J',F',m_F'\right|[P^{(2)}\times I^{(2)}]^{(4)}_m\left|L,J,F,m_F\right\rangle\notag\\
=3*(-1)^{F'-m_F'}\sqrt{(2F+1)(2F'+1)}
\begin{pmatrix}
F'&4&F\\-m_F'&m&m_F
\end{pmatrix}\notag\\
\bullet\begin{Bmatrix}
J'&J&2\\L'&L&2\\F'&F&4
\end{Bmatrix}
\left(L'\right\|P^{(2)}\left\|L\right)\left(J'\right\|I^{(2)}\left\|J\right)\label{e-2-1-2-2}
\end{align}
where the term with $\left(\right)$ is the 3-$j$ symbol, and the term with $\left\{\right\}$ is the 9-$j$ symbol. Then we can transform these matrix elements into a set of basis functions belonging to the irreducible representations of the cubic double group. The eigenfunctions of the Hamiltonian (\ref{e-2-1-2-1}) can be solved by expanding the cubic states in terms of Gaussian functions \cite{Baldereschi1974CCttSMoSAS}. The resulting cubic states are then used as the basis for the dimer and chain models.

\subsection{A pair of acceptors and the one-hole model}\label{pair}

For the case of a pair of acceptors, a calculation for a fully interacting two-hole model has recently been reported in the Heitler-London limit \cite{C.Durst2017HLMfAAIiDS}, but it is challenging to extend this approach to systems with more than two acceptors. We therefore introduce an independent-hole model to simplify the calculation, where we initially assume that there is only one hole in the acceptor pair. The single-hole system can be written as $A^-_2$, where A stands for the acceptor (compare the $H^+_2$ molecule, which contains a single electron). In this case, the Hamiltonian is
\begin{align}
\hat{H}_{s,c}^{\mathrm{pair}}=\hat{H}^A_{s,c}-\frac{2}{r_B}=\hat{H}_{s,c}^B-\frac{2}{r_A}\label{e-2-2-1}
\end{align}
where $H_A$ and $H_B$ are the Hamiltonians of a single acceptor $A$ and a single acceptor $B$ (which may be written either in the spherical approximation or including cubic terms). Then we can easily obtain an expression for the off-diagonal matrix element (or transition strength)
\begin{align}
\left\langle\phi_A|\hat{H}^{\mathrm{pair}}|\phi_B\right\rangle=&\frac{1}{2}(E_A+E_B)\left\langle\phi_A|\phi_B\right\rangle-\left\langle\phi_A|\frac{1}{r_A}|\phi_B\right\rangle\notag\\
&-\left\langle\phi_A|\frac{1}{r_B}|\phi_B\right\rangle\label{e-2-2-2a}\\
\left\langle\phi^{'}_A|\hat{H}^{\mathrm{pair}}|\phi_A\right\rangle=&\left\langle\phi^{'}_A|H_A-\frac{2}{r_B}|\phi_A\right\rangle\notag\\
=&E_A\left\langle\phi^{'}_A|\phi_A\right\rangle-\left\langle\phi^{'}_A|\frac{2}{r_B}|\phi_A\right\rangle\label{e-2-2-2b}
\end{align}
where $E_A$ and $E_B$ are the single acceptor energy states. 

Using Equations (\ref{e-2-2-2a}) and (\ref{e-2-2-2b}), we can obtain the transition strength between any single-acceptor states on any sites. The single-hole energies can then be found by solving a generalised eigenvalue problem provided we can compute the overlap $\left\langle\phi_A|\phi_B\right\rangle$ and the potential term $\left\langle\phi_A|\frac{1}{r_i}|\phi_B\right\rangle$. We follow the methods in the previous paper \cite{Clementi1966ESoLMS} to find the corresponding matrix elements using the Gaussian expansion of the single-acceptor states; Reference-\onlinecite{Clementi1966ESoLMS} gives the result for states up to $P$ orbitals, while the results for higher angular momenta can be obtained by taking further derivatives along the different axis.

This approach becomes exact as (i) the number of single-acceptor states used and (ii) the number of Gaussian functions used to represent each one both tend to infinity.  Since we are interested in the lowest-lying states in silicon, we use only the lowest 4 single-acceptor states ($1S_\frac{3}{2}$, $2S_\frac{3}{2}$, $2P_\frac{3}{2}$, $2P_\frac{5}{2}$ for the spherical case and $1\Gamma^+_8$, $2\Gamma^+_8$, $1\Gamma^+_6$, $1\Gamma^-_8$ for the cubic case) in our basis, as the others are far away from them in energy. For the spherical case the different total angular momenta are mixed in the array but the projection $m_F$, for which the quantisation axis is chosen along the inter-acceptor axis, remains a good quantum number.  For the cubic case, with a general axis direction states of all symmetries are mixed; however time-reversal symmetry guarantees the states still appear in Kramers doublets, which can be thought of as derived from the $m_F=\pm\frac{1}{2}$ and $m_F=\pm\frac{3}{2}$ pairs in the spherical case.

\subsection{Linear acceptor-chain and LCAO model}\label{chain}

\subsubsection{Finite chain}\label{finitechain}

From the one-hole model in \S\ref{pair}, we can develop a similar linear combination of single-acceptor states to describe a finite linear chain of acceptors by similarly adding the potential terms from neighbouring dopants ($V_{potential}$) into the single-acceptor Hamiltonian. 
\begin{align}
\hat{H}_{s,c}^{\mathrm{chain}}=\hat{H}^A_{s,c}-\frac{2}{r_B}+\hat{V}_{\mathrm{potential}}=\hat{H}_{s,c}^B-\frac{2}{r_A}+\hat{V}_{\mathrm{potential}}\label{e-2-3-1}
\end{align}
The details for the transition matrix element are shown in Appendix \ref{transitionelement}. However, the basis states on different acceptors are not orthogonal and hence the overlap matrix $S$ must be included in the construction of the LCAO model. This requires that all the eigenvalues of the overlap matrix must be positive, in order to obtain a well defined generalised eigenvalue problem. Approximations to the overlap matrix, for example truncating it after a finite number of neighbours, may destroy the positive-definiteness of $S$ and make it impossible to solve the eigenvalue problem. This is a problem particularly for small separations, as we will show in \S\ref{finitechainresult}. To minimise this problem, we include in the calculation the influence of the next nearest neighbor by considering the matrix elements between each acceptor and its next nearest neighbor in both the transition matrix and the overlap matrix.

For definiteness we focus on the 10-acceptor finite chain shown in Figure \ref{f-1} (a), where we label the first five acceptors from one end by a, b, c, d, and e. We assume that the separations appear periodically as shown in Figure \ref{f-1} (a), so the chain possesses a dimerisation that can be varied by changing the separations $d_1$ and $d_2$.

\subsubsection{Short-range model}\label{Shortrange}

We refer to the single-hole model including interactions with the next nearest neighbors as the `long-range' model.  This is expected to be a good model for a single hole bound to an array of acceptors and in this case the long-range Coulomb interactions have an important effect on the physics (as shown in \S\ref{finitechainresult}).  However, we may also wish to understand the behaviour of clusters which are at or close to charge neutrality and hence contain many holes (for example, one hole per acceptor), but the motion of the holes is approximately independent of each other.  In that case we expect that the motion of the other holes will effectively screen out these long-range interactions, so we adopt as our approximation to this charge-neutral case a `short-range' model where the effect of the Coulomb potential term in Equation (\ref{e-2-3-1}) is removed.


\subsubsection{Infinite chain}\label{infinitechain}

\begin{figure}
\centering
\includegraphics[scale=0.3]{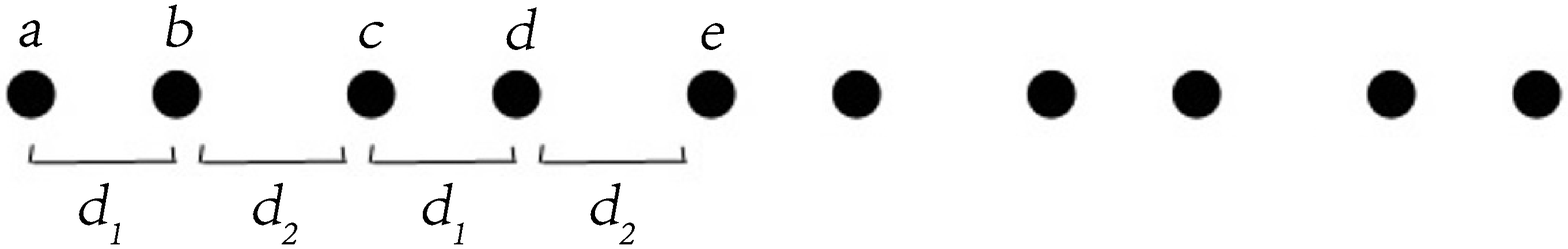}\\
(a)\\
\includegraphics[scale=0.4]{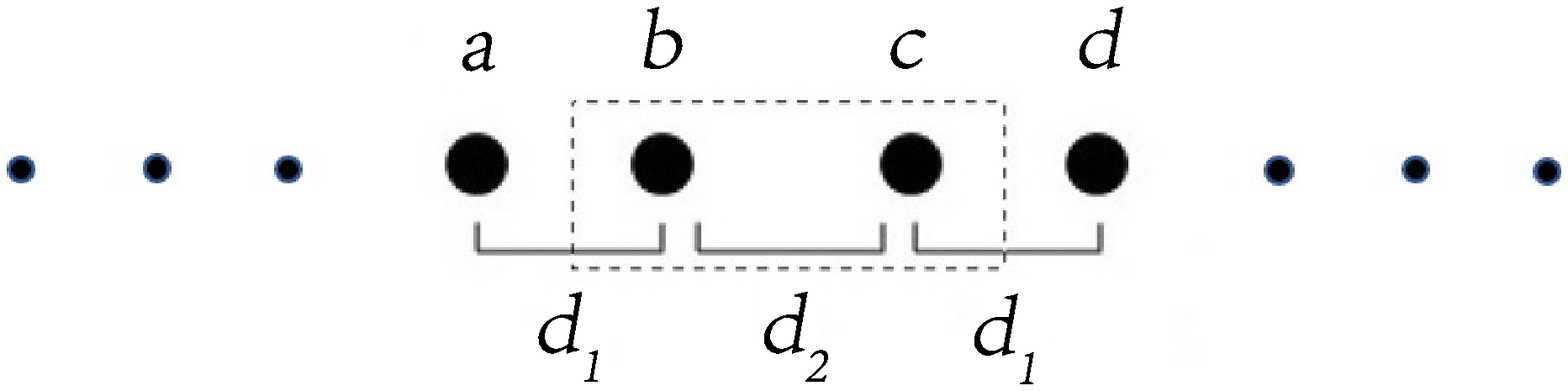}\\
(b)
\caption{(a) Schematic of the 10-acceptor finite chain. (b) The unit cell of the infinite chain; atoms b and c are in the same cell. The letters a, b, c, d and e label the acceptors.  We refer to $d_1<d_2$ as the `short-long' case, and $d_1>d_2$ as the `long-short' case.}\label{f-1}
\end{figure}

From the one-hole model in \S\ref{pair}, we can also generate an LCAO model to describe the linear infinite acceptor-chain in the similar way. The general form of the Hamiltonian will has the same form as for the finite chain (Equation (\ref{e-2-3-1})). We assume each unit cell contains two acceptors as shown in Figure \ref{f-1} (b). The inter-cell separation is taken as $d_1$, and the intra-cell separation is $d_2$. Since the system is periodic its eigenstates are labelled by a Bloch wavevector $k$, which we define so that the phase factor of the transition from left to right is $\mathrm{e}^{\mathrm{i}k}$, and that from right to left is $\mathrm{e}^{-\mathrm{i}k}$.

\subsection{Zak phase}\label{zakphase}

An indication of whether a given state in the finite chain system has a topological origin can be obtained by calculating the Zak phase for the corresponding infinite chain. For a 1D system, this quantity was defined in the previous paper \cite{Delplace2011TZPatEoESiG,Zak1989BPfEBiS} as
\begin{align}
Z=i\int_{first BZ}dk\left\langle u_k|\partial_k u_k\right\rangle\label{e-2-4-1}
\end{align}
where $u_k$ is the eigenvector of the Bloch Hamiltonian at wavevector $k$. When the Zak phase is $0$ modulo $2\pi$, we expect the system to be topologically trivial and the corresponding finite chain to have no topological edge state, whereas when the Zak phase is $\pm\pi$, the system is topologically non-trivial and the corresponding finite chain supports topological edge states. As it is the integration of the Berry connection over the first Brillouin zone, the Zak phase is invariant (modulo $2\pi$) under gauge transformations of the form $\left|u_k\right\rangle\rightarrow e^{i\beta_k}\left|u_k\right\rangle$ \cite{Artacho2017QMiaEHS}.

For a generalized eigenvalue problem, the formula for the Zak phase becomes
\begin{align}
Z=i\int_{first BZ}dk\left\langle u_k|S(k)|\partial_k u_k\right\rangle\label{e-2-4-2}
\end{align}
where $S(k)$ is the overlap matrix.  As previously, Equation (\ref{e-2-4-2}) is invariable under unitary transformations. The correctness of Equation (\ref{e-2-4-2}) will be shown later in \S\ref{infinitechainresult}.

Here we should point out that in our calculations the implied position of the chain end is always at the center of the unit cell.  Hence, different choices of the unit cell for the calculation of the Zak phase correspond to different terminations of the finite chain ---even although these different unit cells describe the same infinite system. So the Zak phase for a particular choice of unit cell actually reflects the properties of the corresponding finite chain as well as the infinite chain, which means the results can be different if we cut the infinite chain in different places. 

\section{Results and discussion}\label{result}

\subsection{Single acceptor problem}\label{singleresult}

\begin{figure}
\centering
\includegraphics[scale=0.3]{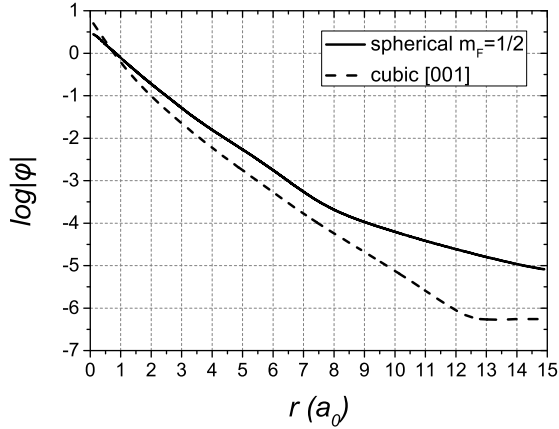}\\
\caption{The behavior of the ground-state wave function for an acceptor in Si, in the spherical case with $m_F=\frac{1}{2}$ (solid line) and the cubic case in the [001] direction (dash line).}\label{f-2}
\end{figure}

The single-acceptor problem can be solved by expanding the wave function of the eigenstates in terms of Gaussian functions \cite{Baldereschi1973SMoSASiS,Baldereschi1974CCttSMoSAS}. We use the lowest four states as a basis for further calculations of the pair and the acceptor chain; their energies in the spherical and cubic cases are shown in Table \ref{t-1}. We note that the states in the cubic case are systematically more strongly bound than those in the spherical case, and we expect they will have correspondingly shorter decay lengths.   This is supported by Figure \ref{f-2},  which shows the behavior of the ground-state wave function for the spherical case with $m_F=\frac{1}{2}$ and for the cubic case in the [001] direction; the more rapid decay in the cubic case is apparent.
\begin{table}
\centering
\caption{The lowest four single acceptor eigenenergies obtained from the Gaussian expansion for Si; the energy unit is the effective Rydberg $R_0$.}\label{t-1}
\begin{tabular}{|c|c|c|c|}
\hline
Spherical state&Spherical result&Cubic state&Cubic result\\
\hline
$1S_\frac{3}{2}$&1.356041&$1\Gamma^+_8$&1.868314\\
\hline
$2P_\frac{3}{2}$&0.456253&$1\Gamma^+_6$&0.930278\\
\hline
$2S_\frac{3}{2}$&0.360829&$1\Gamma^-_8$&0.717426\\
\hline
$2P_\frac{5}{2}$&0.314359&$2\Gamma^+_8$&0.538586\\
\hline
\end{tabular}
\end{table}

\begin{figure*}
\centering
(a)\includegraphics[scale=0.28]{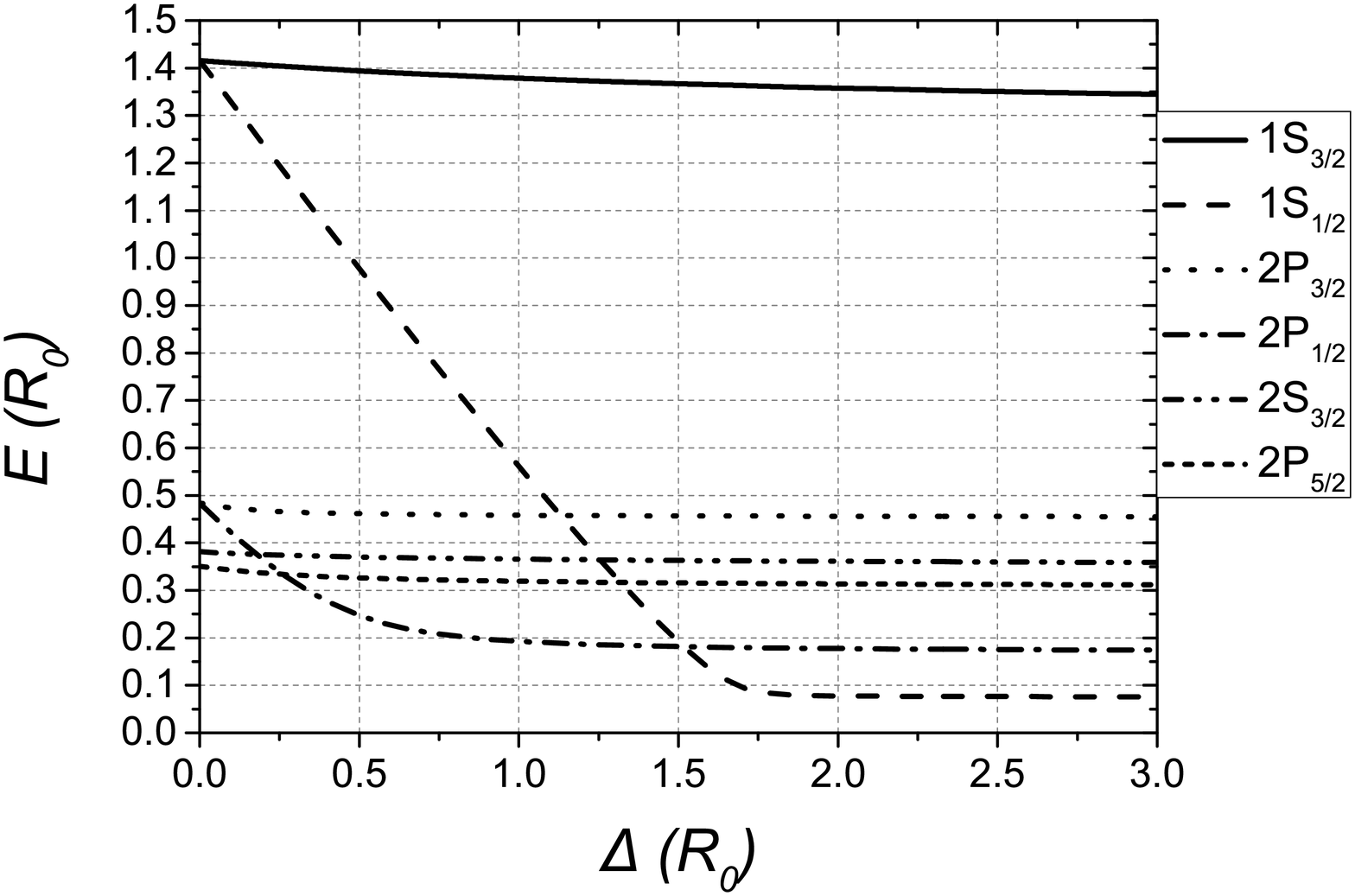}
(b)\includegraphics[scale=0.28]{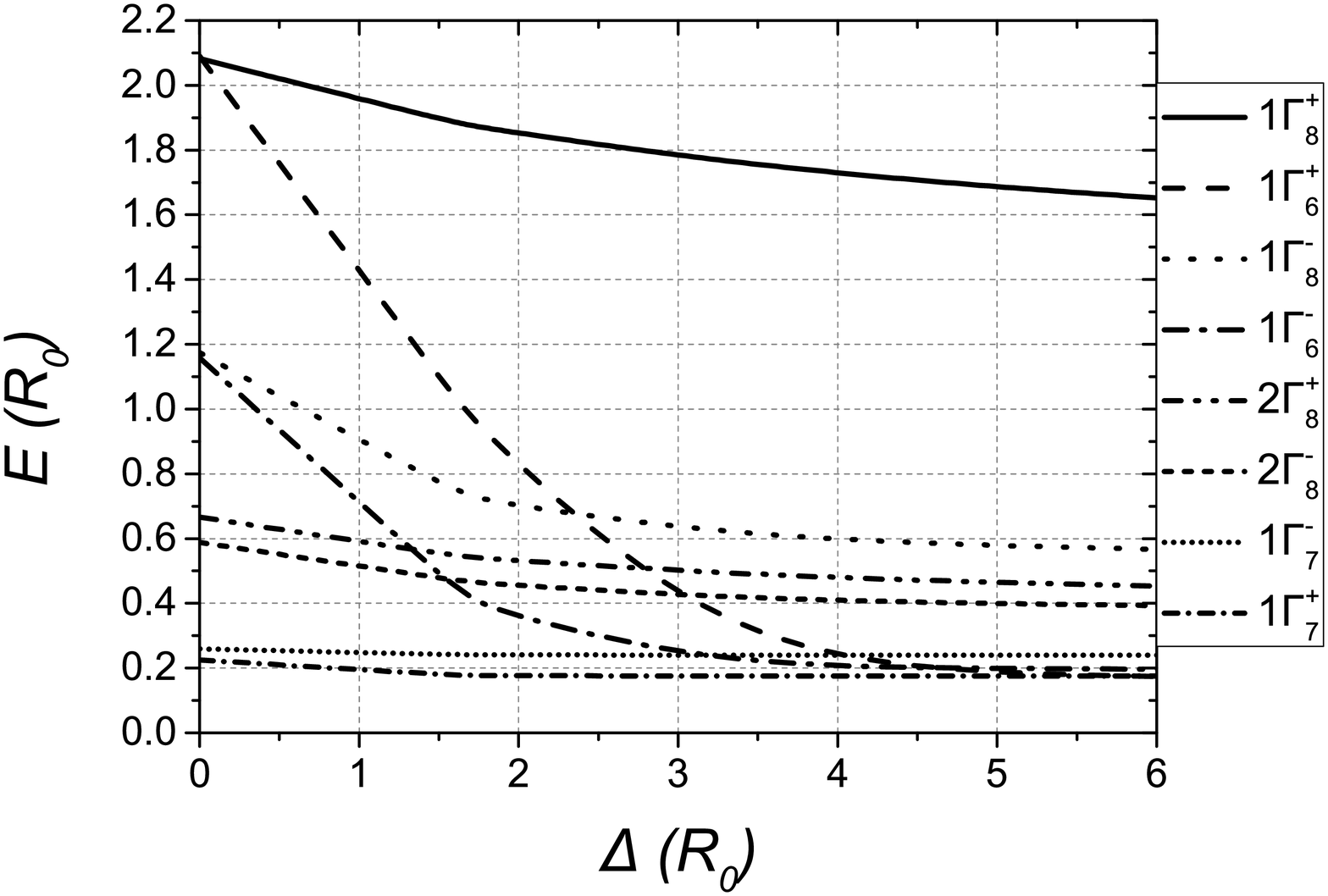}
\caption{The behavior of the single acceptor eigenenergies as a function of $\Delta$ with $\mu=0.483$ and $\delta=0.249$ (the values in Si): (a) the energy spectra for the spherical case, (b) the energy spectra for the cubic case.  Note that some states of other symmetries are not shown, and $\Delta=1.774R_0$ for silicon.}\label{f-3}
\end{figure*}

We also show the behavior of the eigenstates in the spherical and cubic cases as the spin-orbit coupling $\Delta$ changes, for fixed $\mu$ and $\delta$ ($\mu=0.483$, $\delta=0.249$), in Figure \ref{f-3}.  As $\Delta\rightarrow0$, $\vec{F}=\vec{L}+\vec{I}+\vec{S}$ is not the only conserved quantity; instead, $\vec{S}$ and $\vec{L}+\vec{I}$ are separately conserved.  So, the $1S_\frac{3}{2}$ and $1S_\frac{1}{2}$ states converge to the $1S_1$ state of the orbital-only model (where the suffix now refers to the value of $\vec{L}+\vec{I}$); similarly $2S_\frac{3}{2}$ converges to the $2S_1$ state, $2P_\frac{1}{2}$ and $2P_\frac{3}{2}$ will converge to the $2P_1$ state, and $2P_\frac{5}{2}$ will converge to the $2P_2$ state; the $1S_1$, $2S_1$, $2P_1$ and $2P_2$ states were discussed for weak spin-orbit coupling in the early paper \cite{Baldereschi1973SMoSASiS}.  Similarly, in the cubic case without spin-orbit coupling, the symmetries reduce to $\Gamma^\pm_n\otimes \Gamma^+_6$ (where $\Gamma^\pm_n$ denotes the symmetry of the orbital part, including the envelope function and the orbital angular momentum of the atomic p states, and $\Gamma^+_6$ is the symmetry of a single spin-1/2).

\subsection{A pair of acceptors with one hole}\label{pairresult}

The behaviors of the eigenenergies for the spherical and cubic cases are shown as a function of acceptor separation $r$ in Figure \ref{f-4}. The closely related case of the $H^+_2$ molecule is discussed in Appendix \ref{h2+} (and shown in Figure \ref{f-10}).   The states all converge to one of the lowest four states of a single acceptor as $r\rightarrow\infty$, and can roughly be understood as either bonding or antibonding combinations of the single-acceptor states; however, for the cubic cases this is complicated by crossings of the states. The splittings between the states set in at smaller values of $r$ for the cubic case compared to the others.
\begin{figure*}
\centering
(a)\includegraphics[scale=0.28]{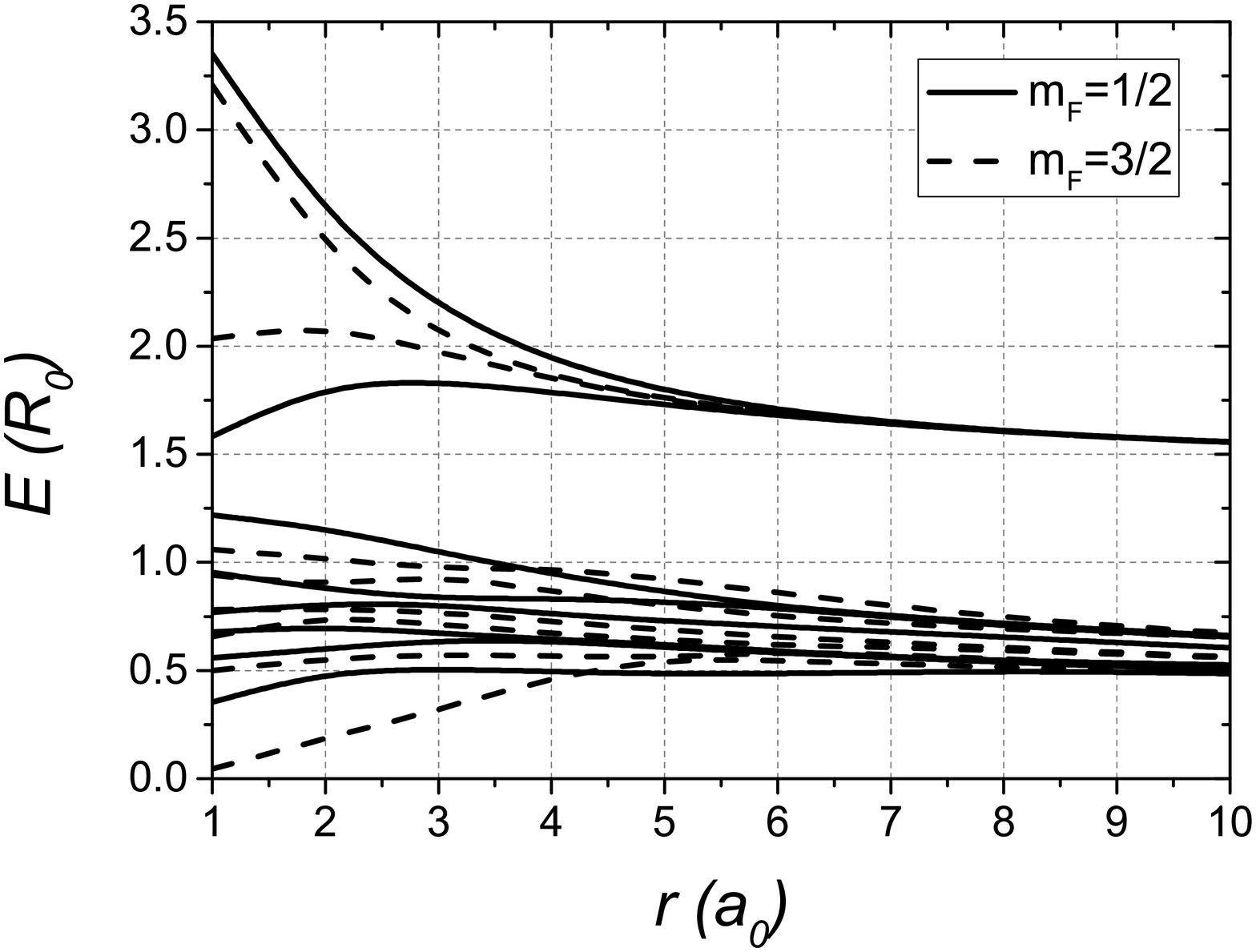}
(b)\includegraphics[scale=0.28]{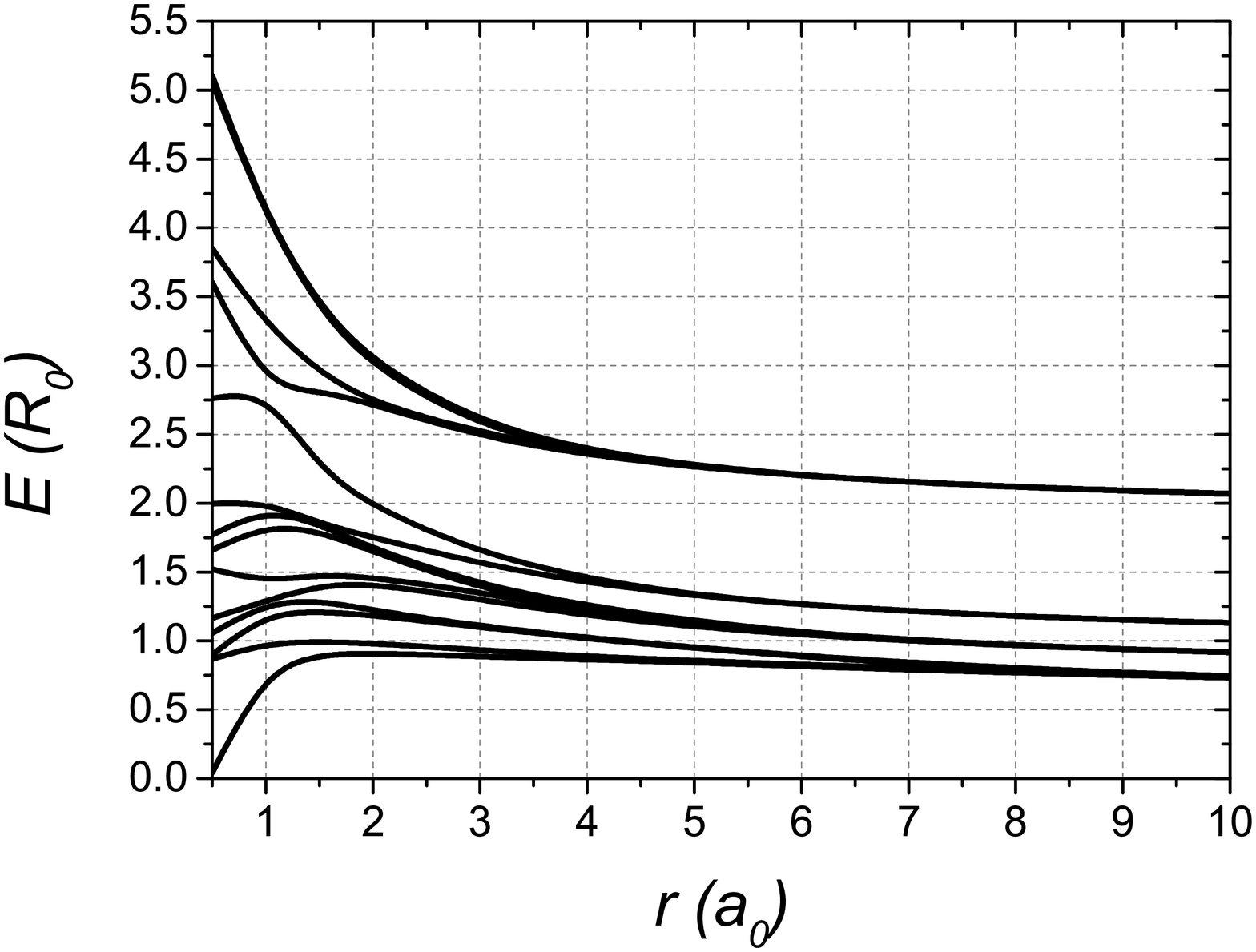}
(c)\includegraphics[scale=0.28]{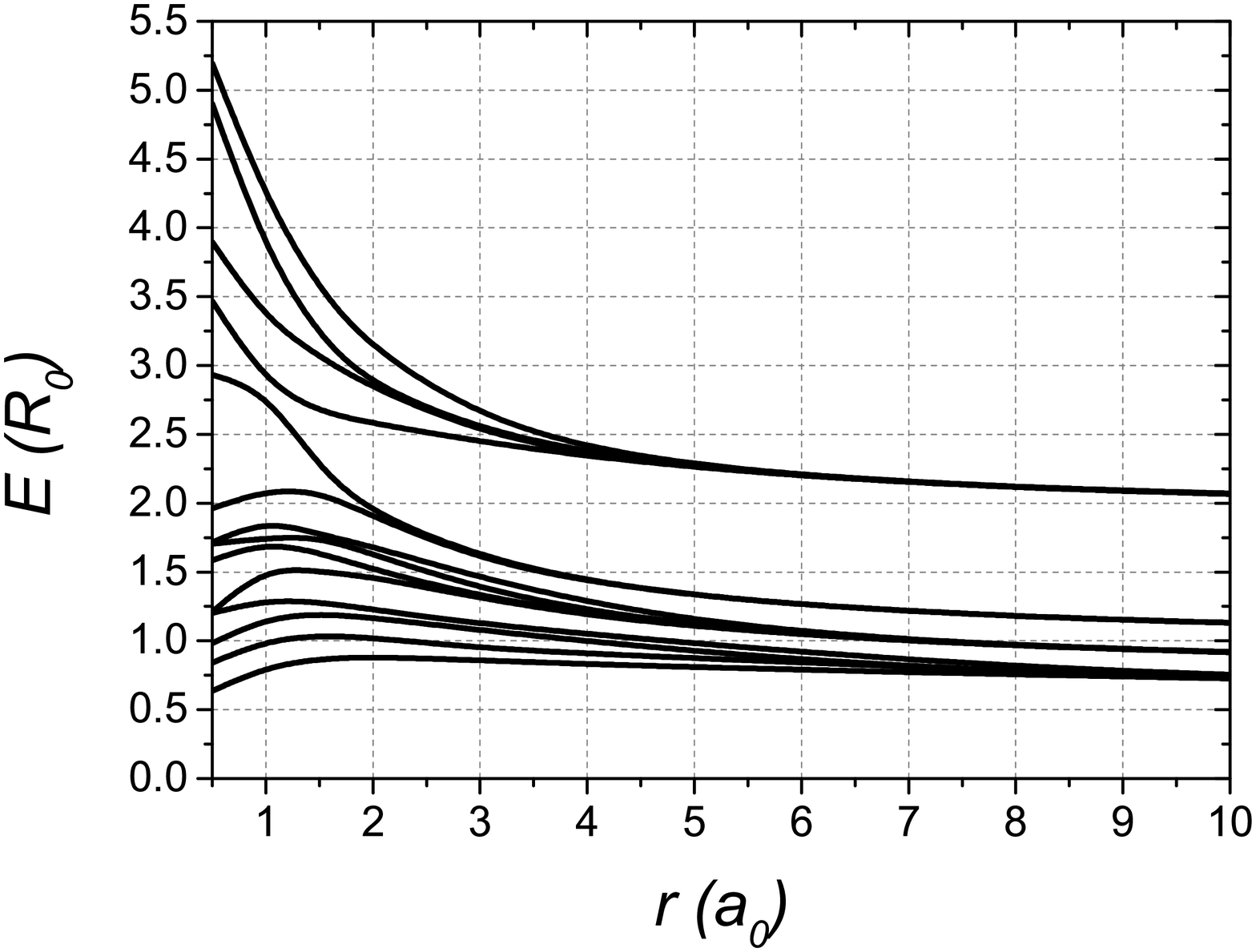}
(d)\includegraphics[scale=0.28]{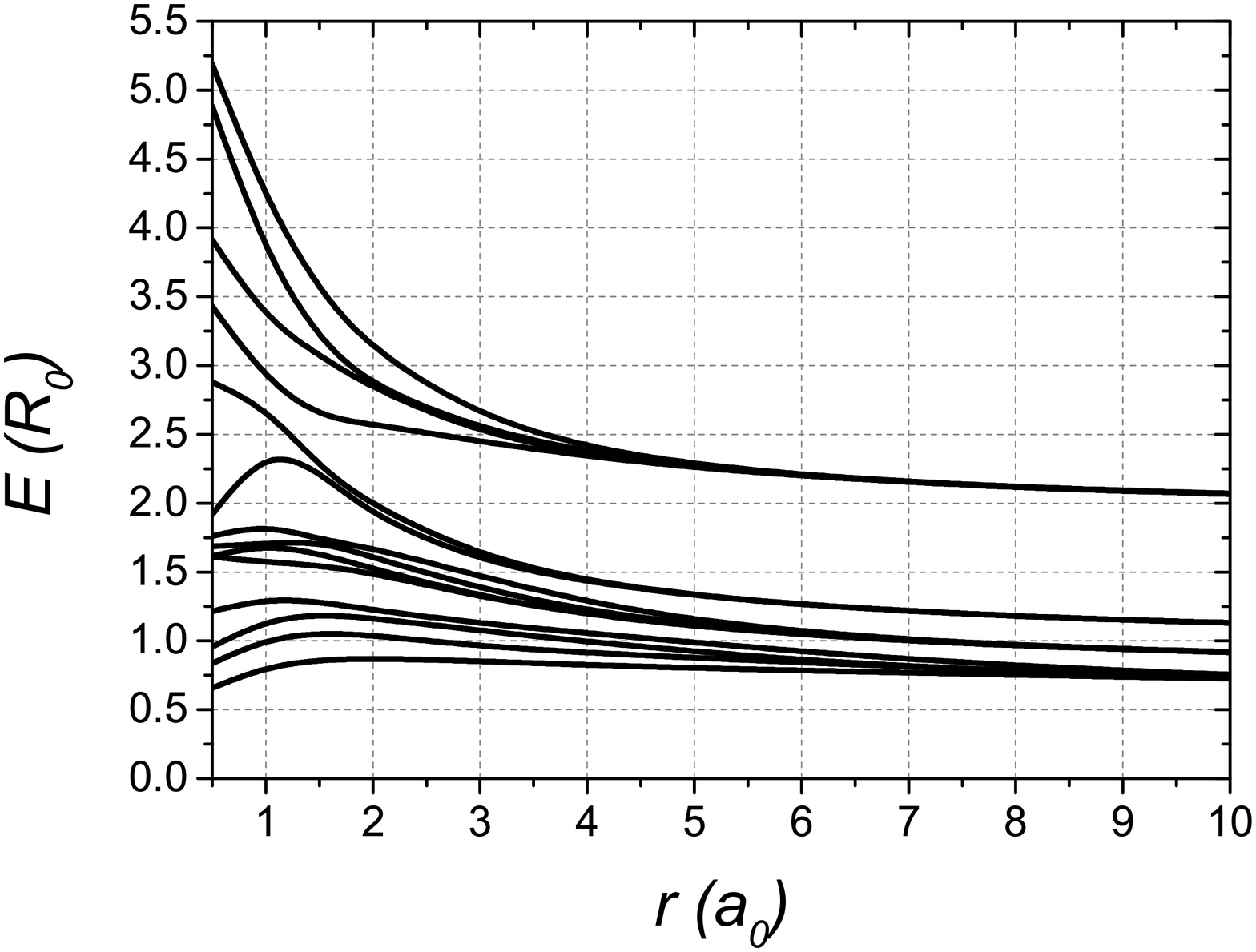}
\caption{Eigenenergies for a pair of acceptors with one hole as a function of separation $r$ for (a) the spherical case, and for the cubic case in (b) the [001] direction,  (c) the [110] direction,  (d) the [111] direction. In (a), the states with $m_F= 3/2$ ($m_F= 1/2$) are the dash lines (solid lines).}\label{f-4}
\end{figure*}

\subsection{Linear acceptor-chain}\label{chainresult}

\subsubsection{Finite chain}\label{finitechainresult}

As our LCAO model does not contain the influence of all the acceptors in the chain, the overlap matrix $S$ is not guaranteed to be positive definite. For example, the behavior of the eigenvalues of the overlap matrix for a 10-acceptor finite chain in the spherical model (truncated at the next-nearest-neighbor) when $m_F=\frac{1}{2}$ as a function of the acceptor separation $d_2$ when $d_1=4a_0$ is shown in Figure \ref{f-5} in Appendix \ref{overlapeigenvalues}. It can be seen that, even with the influence of the next-nearest-neighbor included, $S$ ceases to be positive definite for separations $d_2<6a_0$.  The next-next nearest term and the following terms are small compared to the next-nearest-neighbor ones, so adding them only improves the description of the system a little but significantly increases the cost of the calculation. Therefore, we only include the next-nearest-neighbor terms in our model, and restrict our calculations so we do not enter the parameter regions where the corresponding $S$ matrix is not positive definite. For the spherical case we require that one of the separations is larger than $4a_0$ and  the other is no smaller than $6a_0$; for the cubic case (where the basis states are more localised) we require that one of the separations is larger than $2a_0$ and the other is no smaller than $4a_0$. From now on, we refer to the case  $d_1<d_2$ as the short-long arrangement, to $d_1=d_2$ as the uniform chain, and to $d_1>d_2$ as the long-short arrangement.

\begin{figure*}
\centering
(a)\includegraphics[scale=0.28]{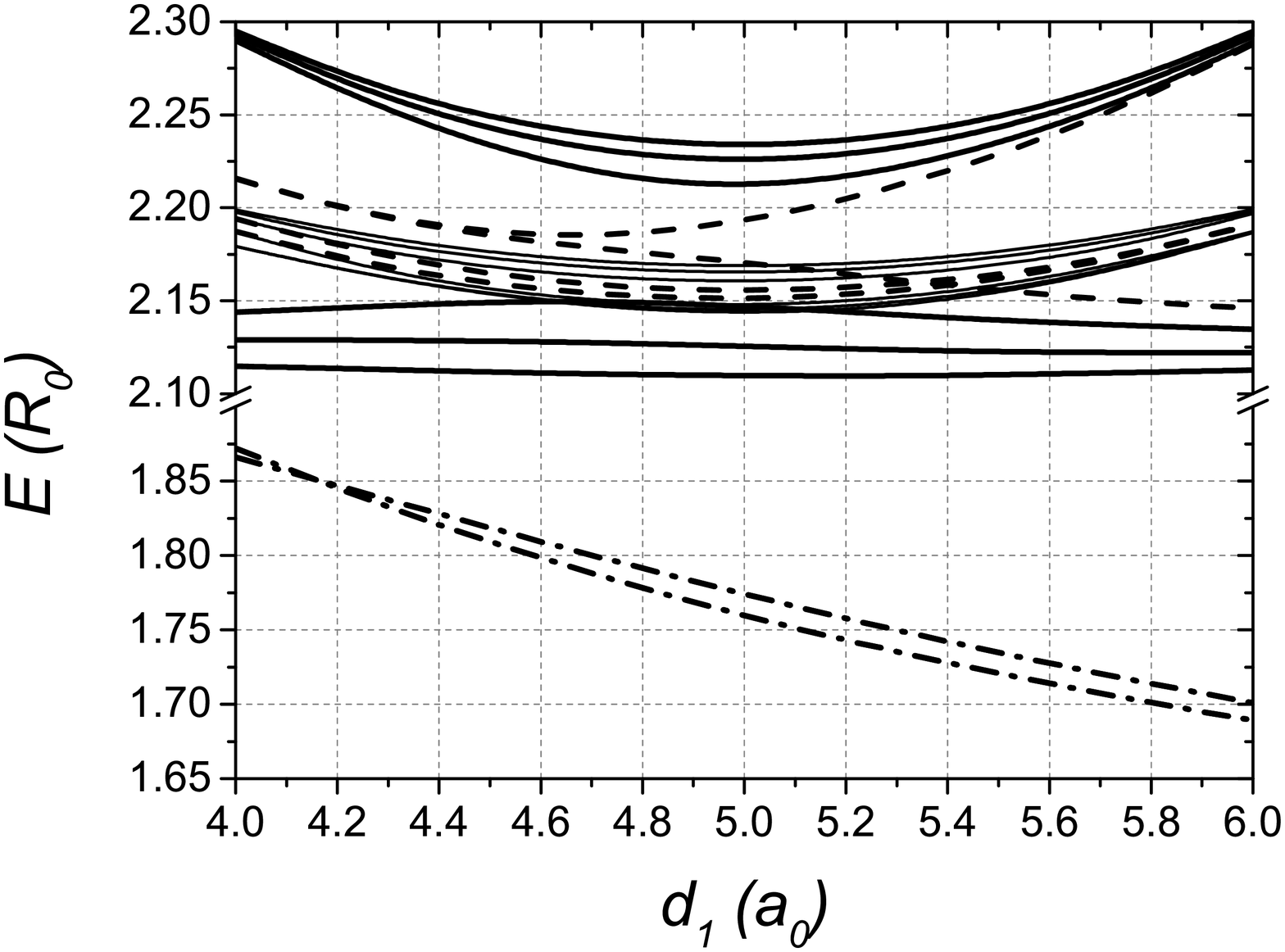}
(b)\includegraphics[scale=0.28]{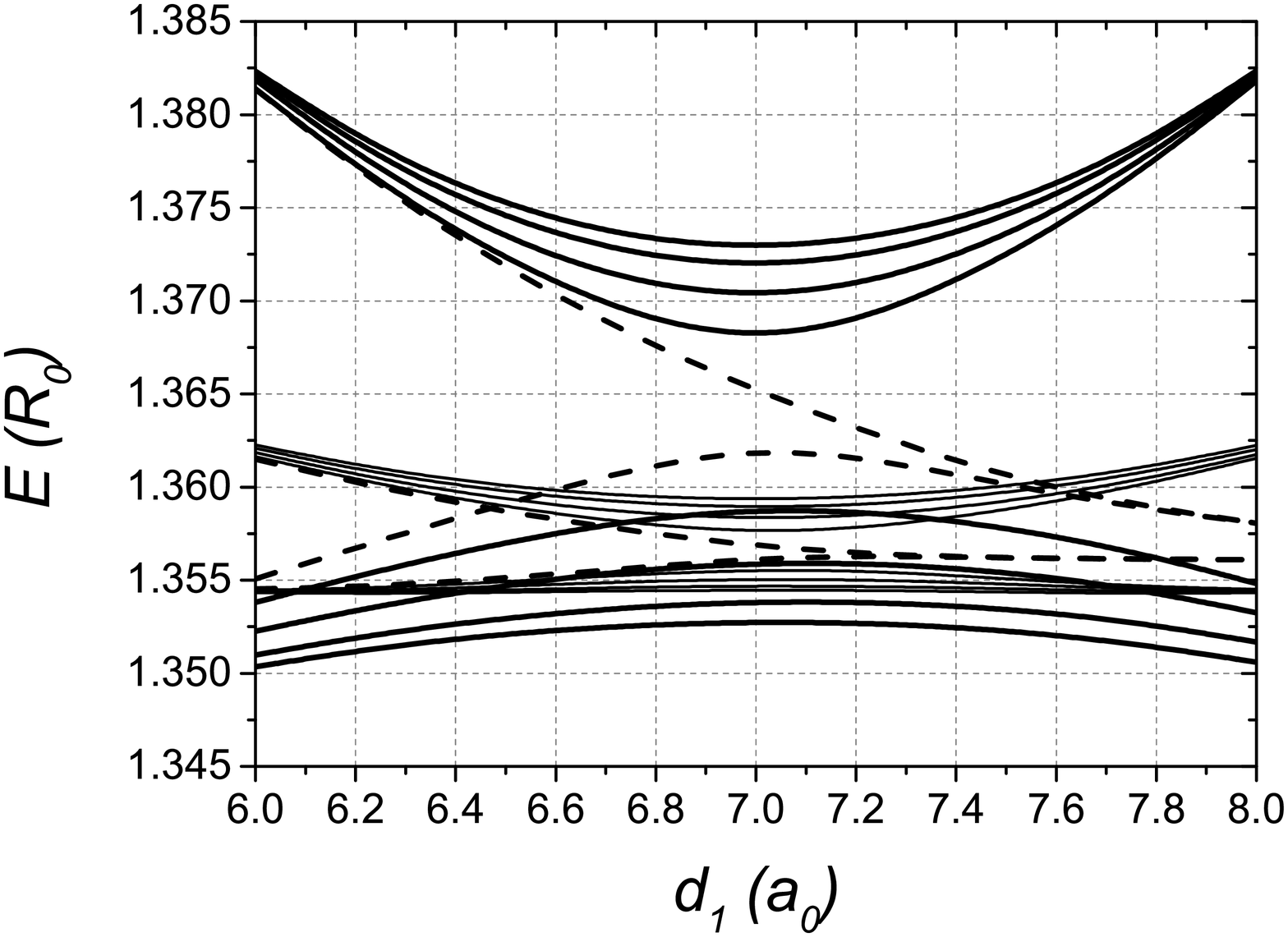}
(c)\includegraphics[scale=0.28]{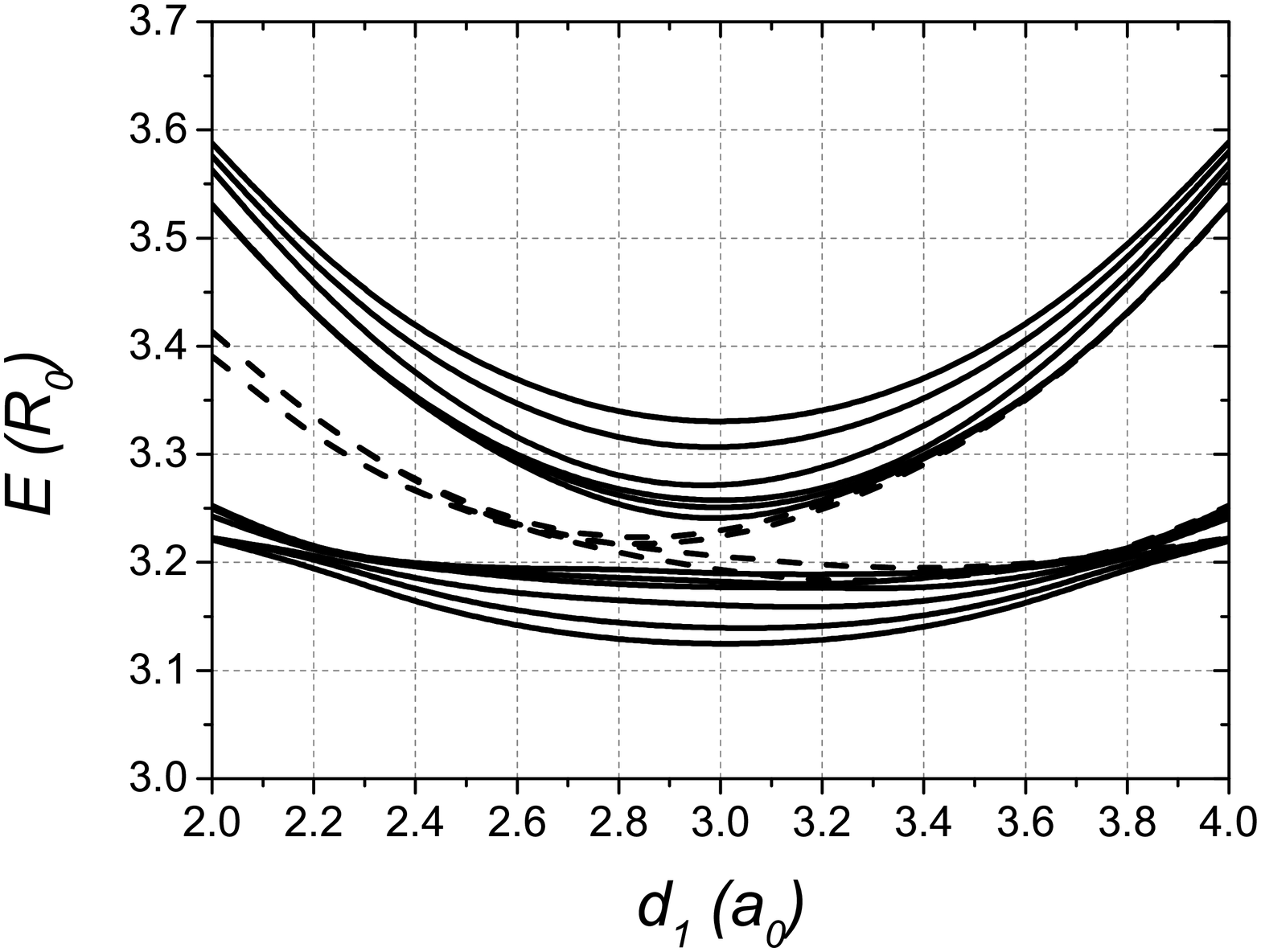}
(d)\includegraphics[scale=0.28]{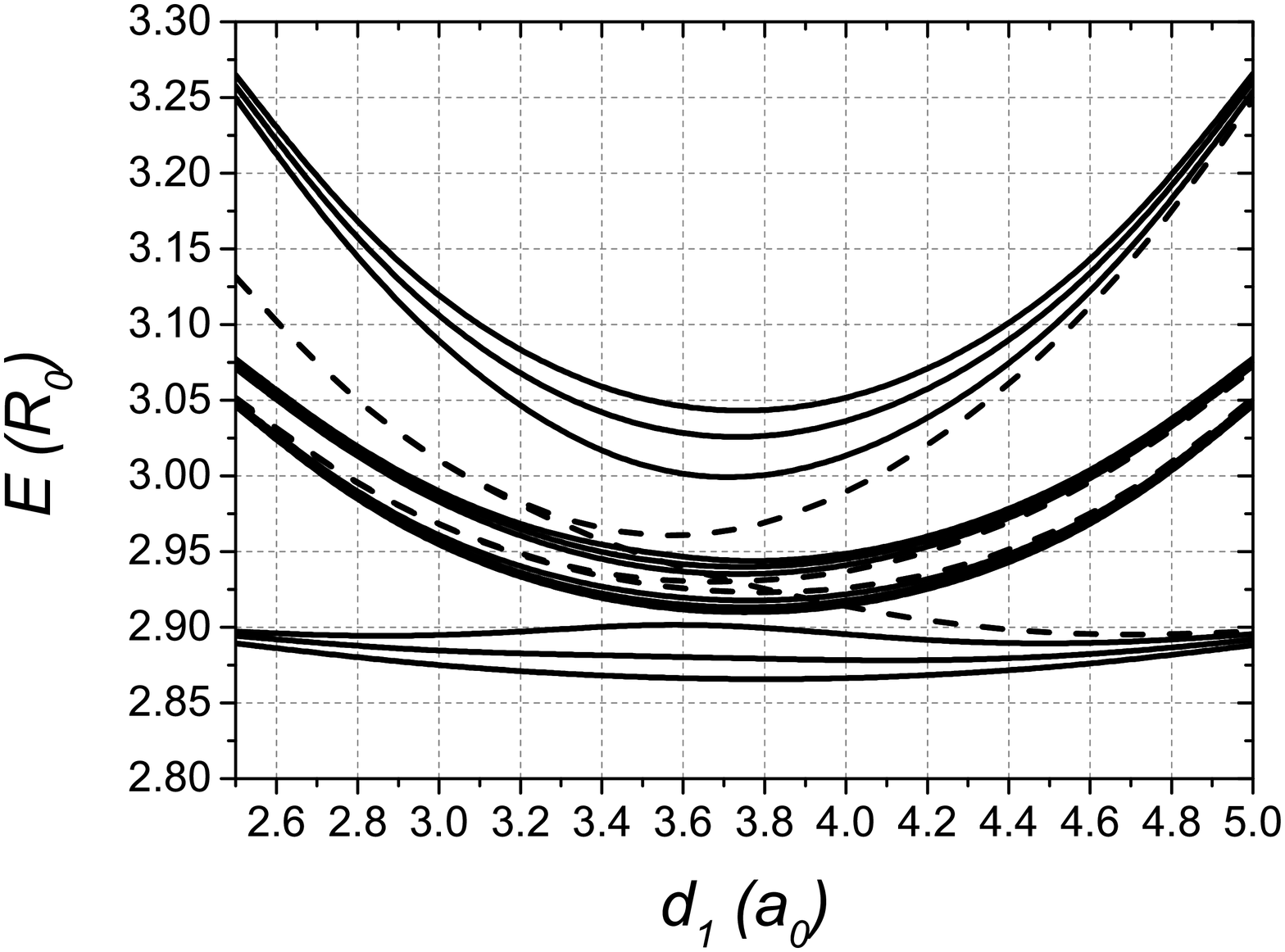}
(e)\includegraphics[scale=0.28]{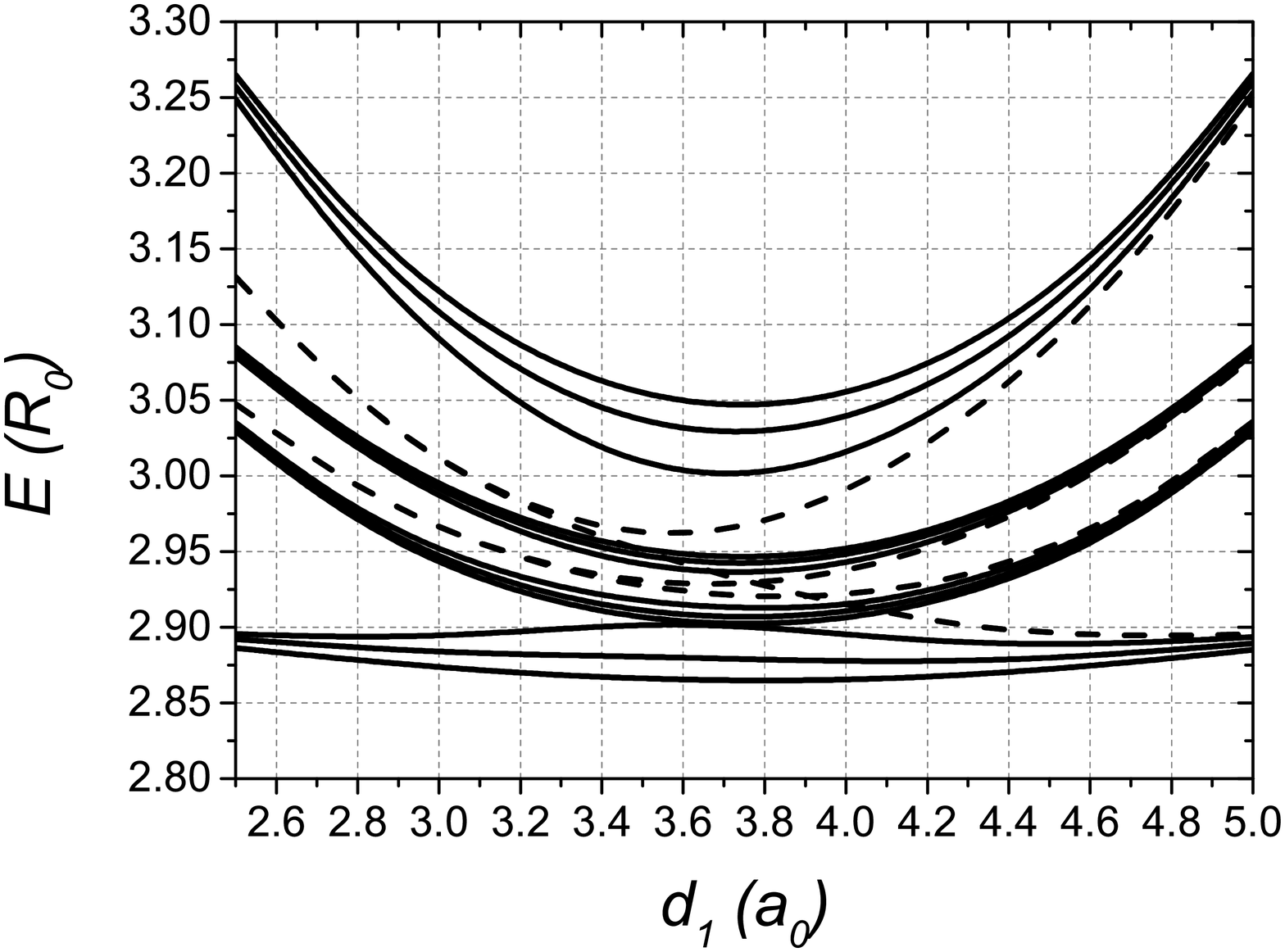}
\caption{The lowest few energy states for the finite chain as a function of $d_1$ when $d_1+d_2$ is held constant. (a) the lowest 20 energy states for the spherical case when $d_1+d_2=10a_0$, (b) the lowest 20 energy states for the spherical case in the short-range model without long-range potential when $d_1+d_2=14a_0$, (c) the lowest 32 states for the cubic case in the [001] direction when $d_1+d_2=6a_0$, (d) the lowest 32 energy states for the cubic case in the [110] direction when $d_1+d_2=7.5a_0$, (e) the lowest 32 energy states for the cubic case in the [111] direction when $d_1+d_2=7.5a_0$. The dash lines are the states splitting from the main bands and lying between them. For the spherical case shown in (a), the dash-dot lines are the states that split from the main bands and lie below them; the solid black lines show states in the main bands for $m_F=\frac{1}{2}$, the solid gray lines are the main bands for $m_F=\frac{3}{2}$. The same style-coding was also done for (b).}\label{f-6}
\end{figure*}

First, we fix the sum $d_1+d_2$ to a constant, choosing the values $10a_0$ for the spherical model, $6a_0$ for the cubic model in the [001] direction, and $7.5a_0$ for the cubic model in the [110] and [111] direction (This is because the overlap matrix for the infinite chain is not positive definite in the [110] and [111] directions under the condition $d_1+d_2=6a_0$ --- see \S\ref{infinitechainresult}). The behavior of the lowest few energy states as a function of $d_1$ under this condition is shown in Figure \ref{f-6}; we show the lowest 20 energy states for the spherical case, and the lowest 32 states for the cubic case. It can be seen that the eigenstates are arranged in groups that correspond to the bands in the infinite-chain model (see below).  For the spherical case in Figure \ref{f-6}(a), the bulk states with $m_F=1/2$ are shown in solid black lines and those with $m_F=3/2$ in solid gray lines, while pairs of states shown in dash lines split off from these main bands.  In each case a nearly degenerate pair (dash line) lies in the gap between main bands on one side of the diagram, and converges into two different bands on the other side. These states in dash lines are each topologically non-trivial on one side of the diagram. There are also other states that always lie below the main bands (the dash-dot lines). We find that the dash-dot lines below the main band are localized at the end of the chain, and the dash lines between the main bands are localized on the acceptors next to the end of the chain.  (The dash-dot lines are not shown for the cubic cases --- we explain the reason below.)

Now let us investigate the electrostatic origin of the edge states below the main band. These states are introduced into our system because of the parabolic potential arising from the long-range interactions between the charges. This potential rises at the ends of the chain, reflecting the different environments of the acceptors in the middle and at the ends of the chain, so when we add a hole to either of the highest two states among them, they will be localized at the ends. We can check the influence of the parabolic potential by comparing the results for the short-range model, where the long-range Coulomb interactions are absent.  Without the long-range potential, the system is less localized than the original one, so we can only have $d_1+d_2=14a_0$ while retaining a positive definite overlap matrix. The behavior of the lowest 20 energy states as a function of $d_1$ under this condition is shown in Figure \ref{f-6} (b). We see that the dash-dot lines below the main band disappear (This is also true for the short-range model in the cubic case, not shown). Since these edge states arise purely from electrostatic effects they are trivial (i.e. non-topological) states, and we do not show them in the graphs for the cubic cases.  

Comparing Figure \ref{f-6}(a) and (b), we also see that the behavior of the dash line edge states associated with the $m_F=1/2$ (solid black line) bands in the spherical model reverses: for the long-range model (a) the states lie in the band gap for the short-long arrangement but join the bands in the long-short case, while the reverse is true in the short-range model (b).  This is because the long-range electrostatic interactions effectively pull the end acceptors away from the bulk bands, transforming a chain ending with a long bond into one ending with a short bond and \textit{vice versa}.  This is also reflected in the different numbers of (solid black line) band states with $m_F=1/2$ in the two cases. 

We can also see that the behavior of the dash line edge states associated with the $m_F=3/2$ (solid gray line) bands in the spherical model does \textit{not} reverse between the long-range and short-range cases, even though the number of states in each band changes just as for $m_F=1/2$ as the electrostatic edge state is pushed back into the band.  As we show in \S\ref{infinitechainresult}, this is a consequence of an anomalous variation of the effective transition amplitude with distance in the particular geometry considered; it is related to an anomalous behaviour of the topological Zak phase that is discussed in \S\ref{infinitechainresult}. 

The calculations for the cubic cases (Figure \ref{f-6} (c) to (e)) are performed in the long-range model, and the behaviour of the edge states (dash line) is similar to the long-range spherical model. For the [001] and [111] directions, the $m_F=\pm 3/2$ and $m_F=\pm1/2$ bands of the spherical model evolve into states which retain different symmetries in the cubic environment; a dash line can therefore cross all the states in a band having a different symmetry from its own.  In the [110] direction, on the other hand, there is just one irreducible representation that is even under exchange of the acceptors and one that is odd, so a given dash line will anti-cross (with states of the same symmetry) or cross (with states of the opposite symmetry) alternately as it passes through a band; we nevertheless make the dash line continuously as if it crossed all the other states (the anti-crossings are hardly visible on the scale of Figure \ref{f-6}(d)). The relevant symmetries are shown in Table \ref{t-3}.

\begin{table}
\centering
\caption{The symmetry for the ground state under the cubic model.}\label{t-3}
\begin{tabular}{|c|c|c|}
\hline
System&Group&Symmetry\\
\hline
Single Acceptor&$O_h$&$\Gamma^+_8$\\
\hline
Pair/Chain([001] direction)&$D_{4h}$&$\Gamma^\pm_6,\Gamma^\pm_7$\\
\hline
Pair/Chain([110] direction)&$D_{2h}$&$\Gamma^\pm_5$\\
\hline
Pair/Chain([111] direction)&$D_{3d}$&$\Gamma^\pm_4,\Gamma^\pm_5,\Gamma^\pm_6$\\
\hline
\end{tabular}
\end{table}


\subsubsection{Infinite chain}\label{infinitechainresult}

\begin{figure*}
\centering
(a)\includegraphics[scale=0.26]{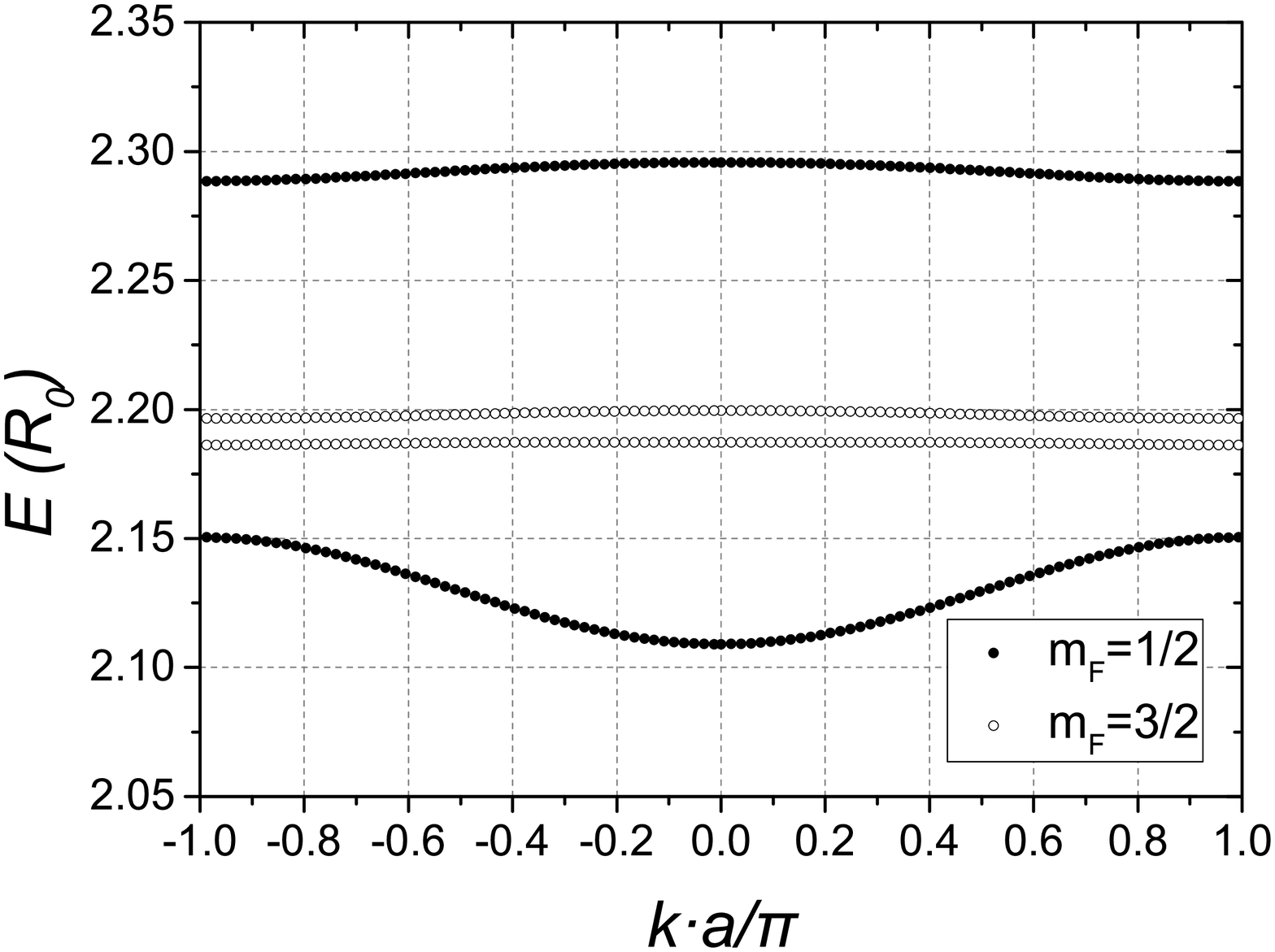}
(b)\includegraphics[scale=0.26]{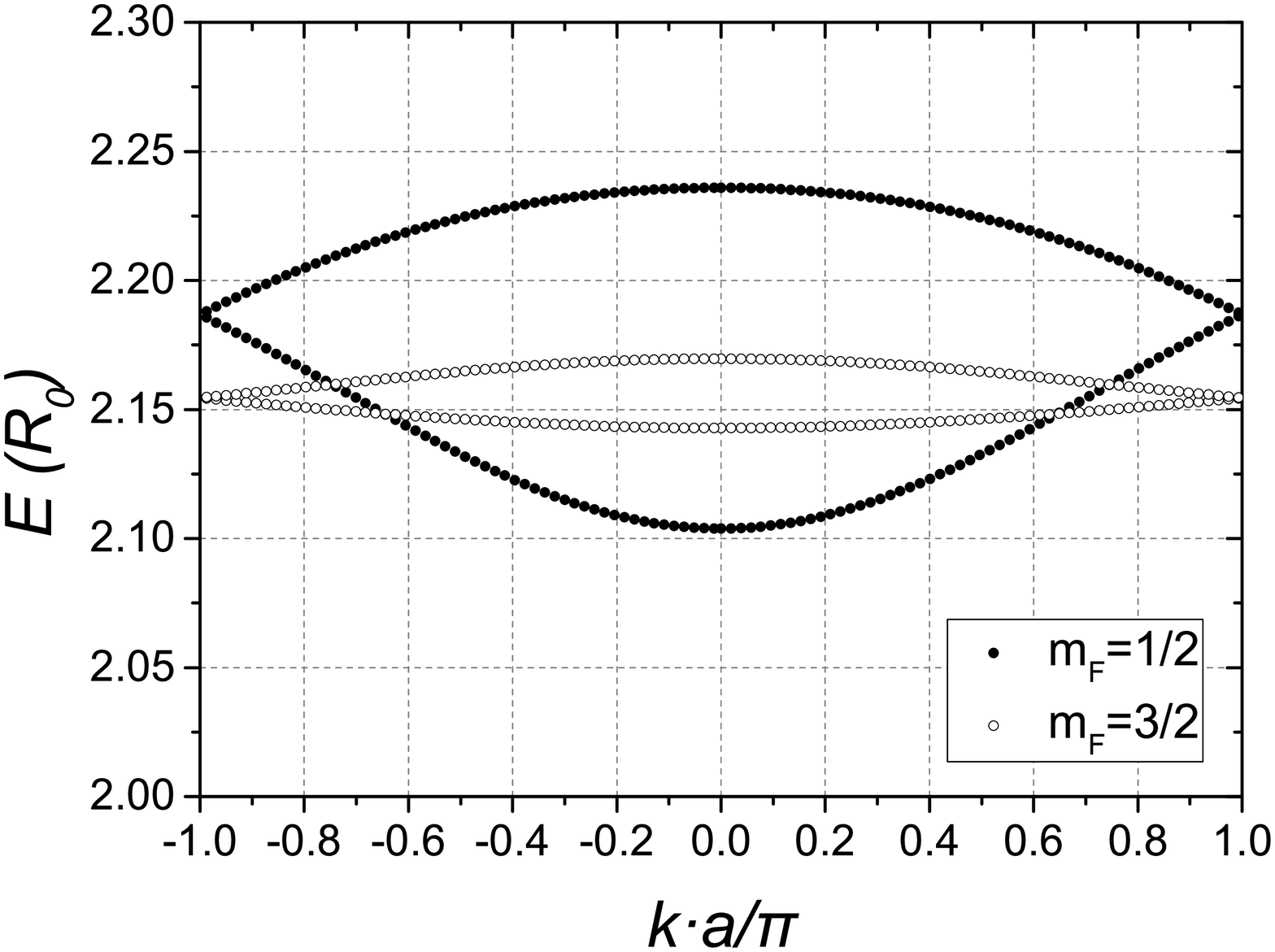}
(c)\includegraphics[scale=0.26]{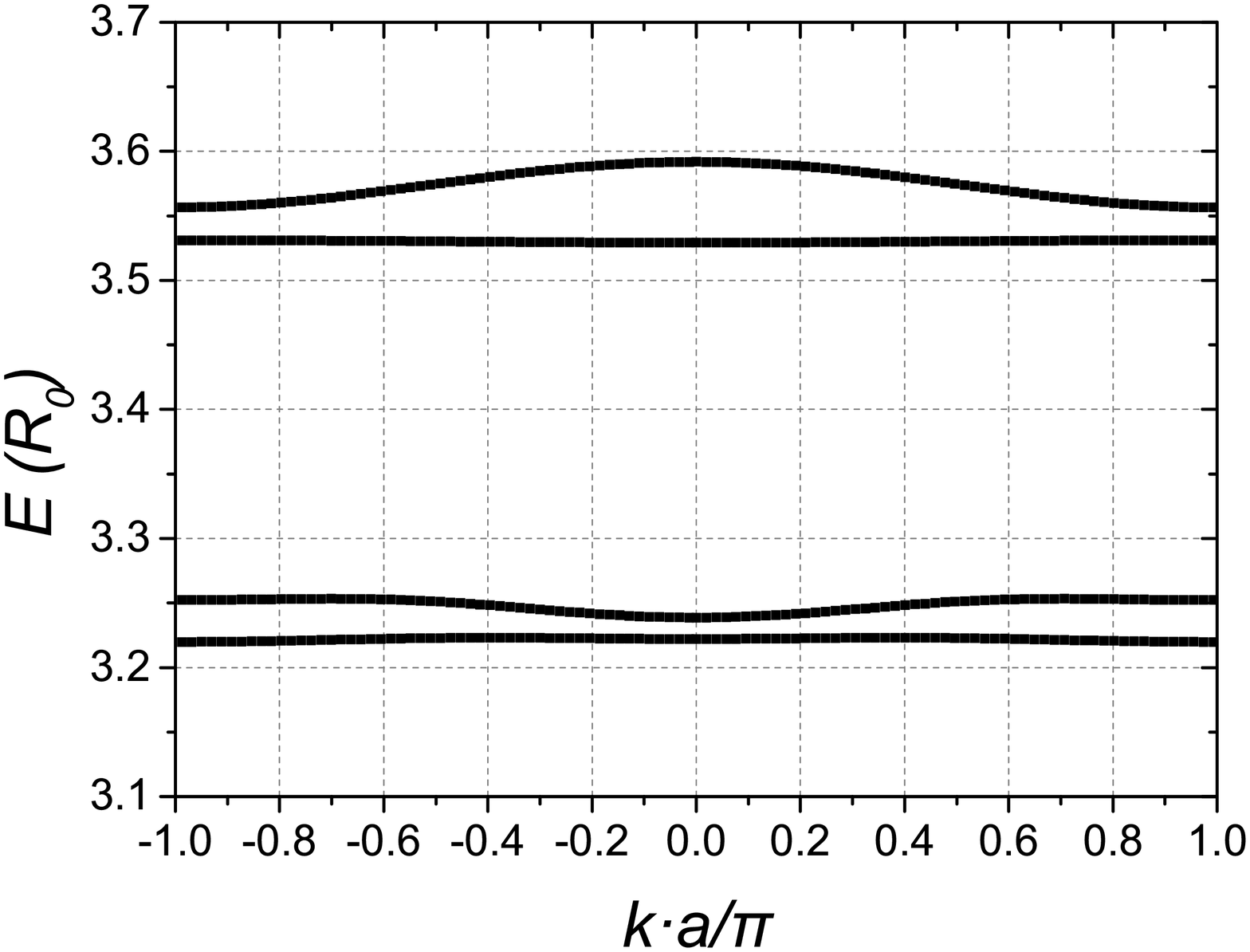}
(d)\includegraphics[scale=0.26]{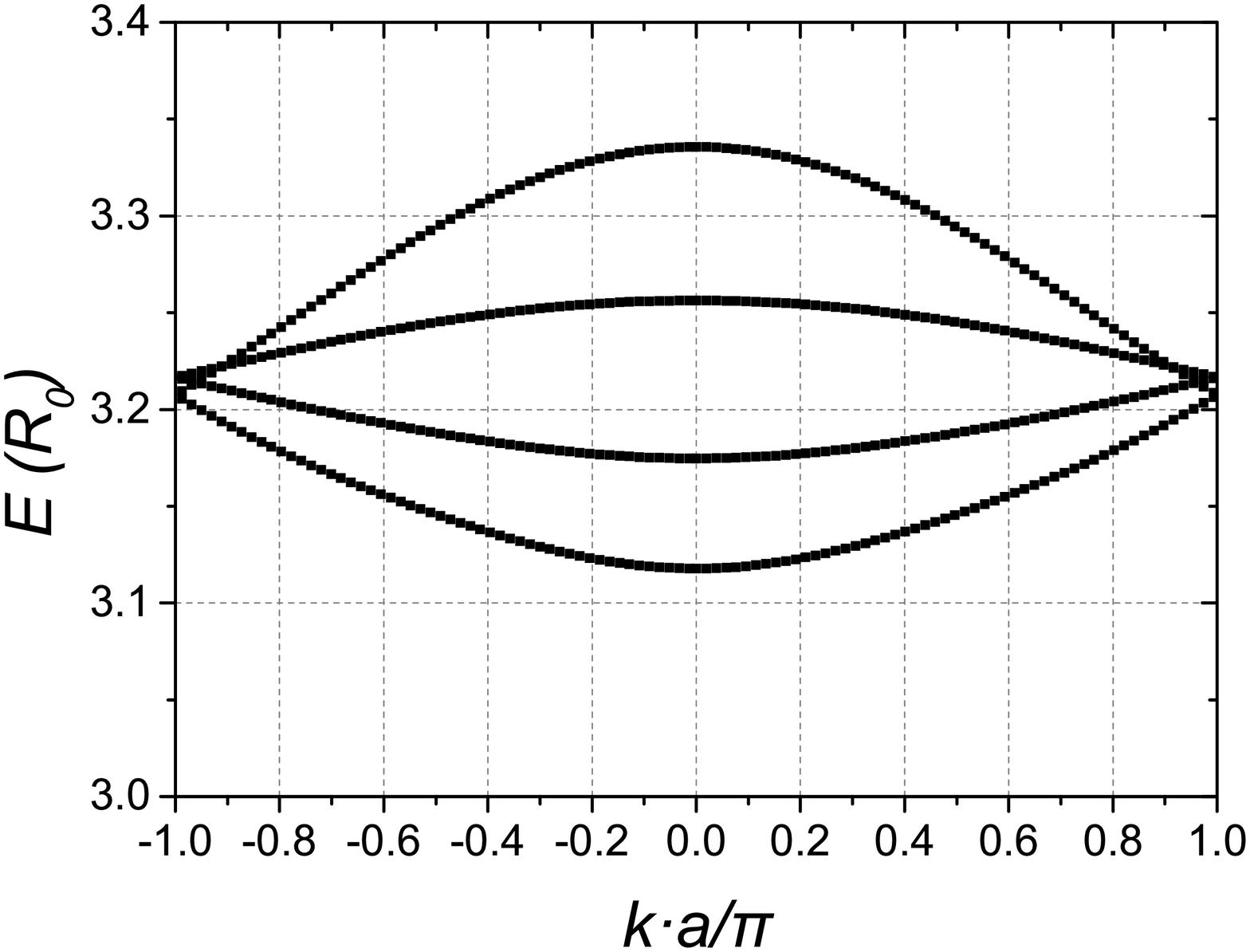}
(e)\includegraphics[scale=0.26]{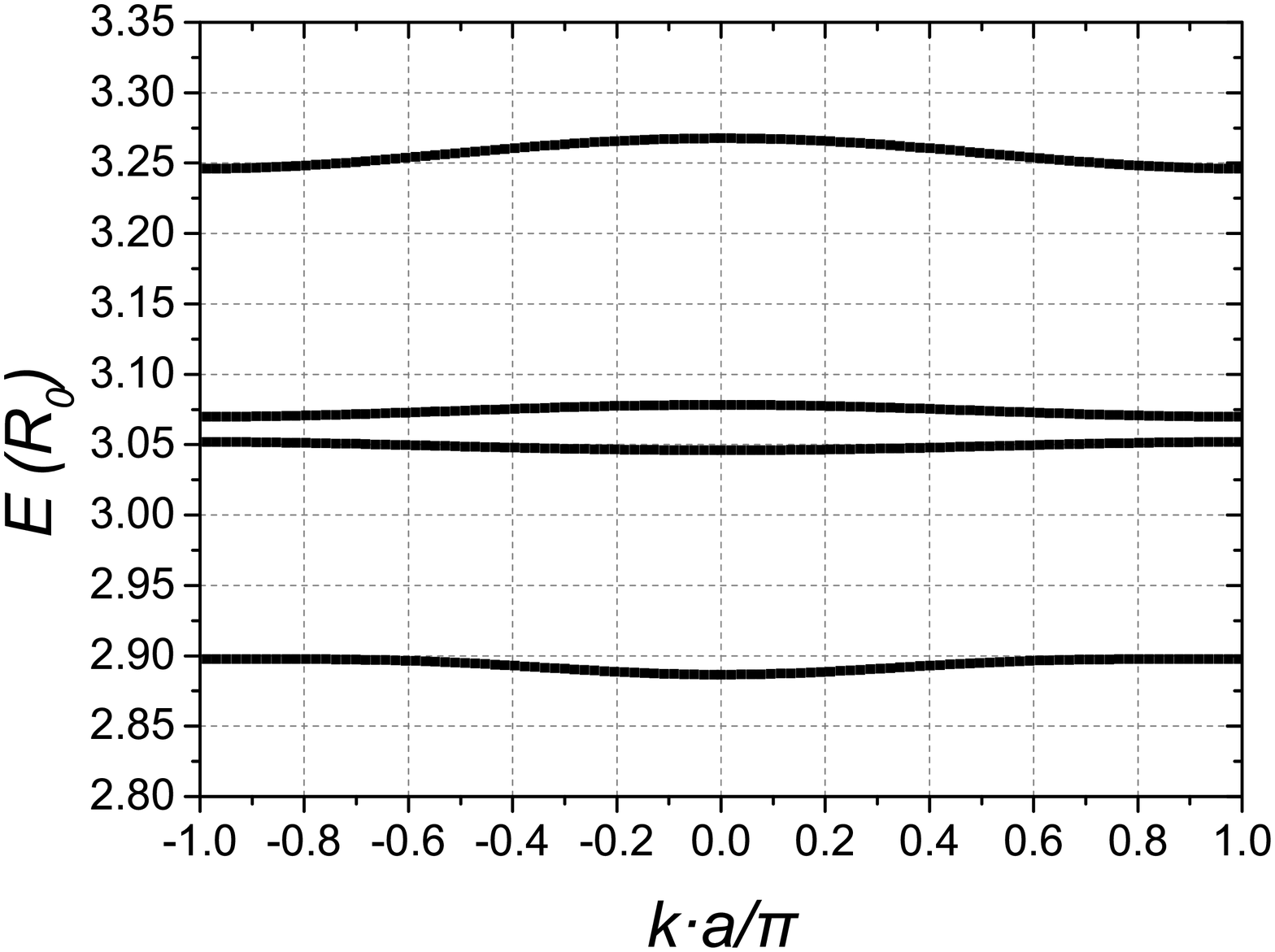}
(f)\includegraphics[scale=0.26]{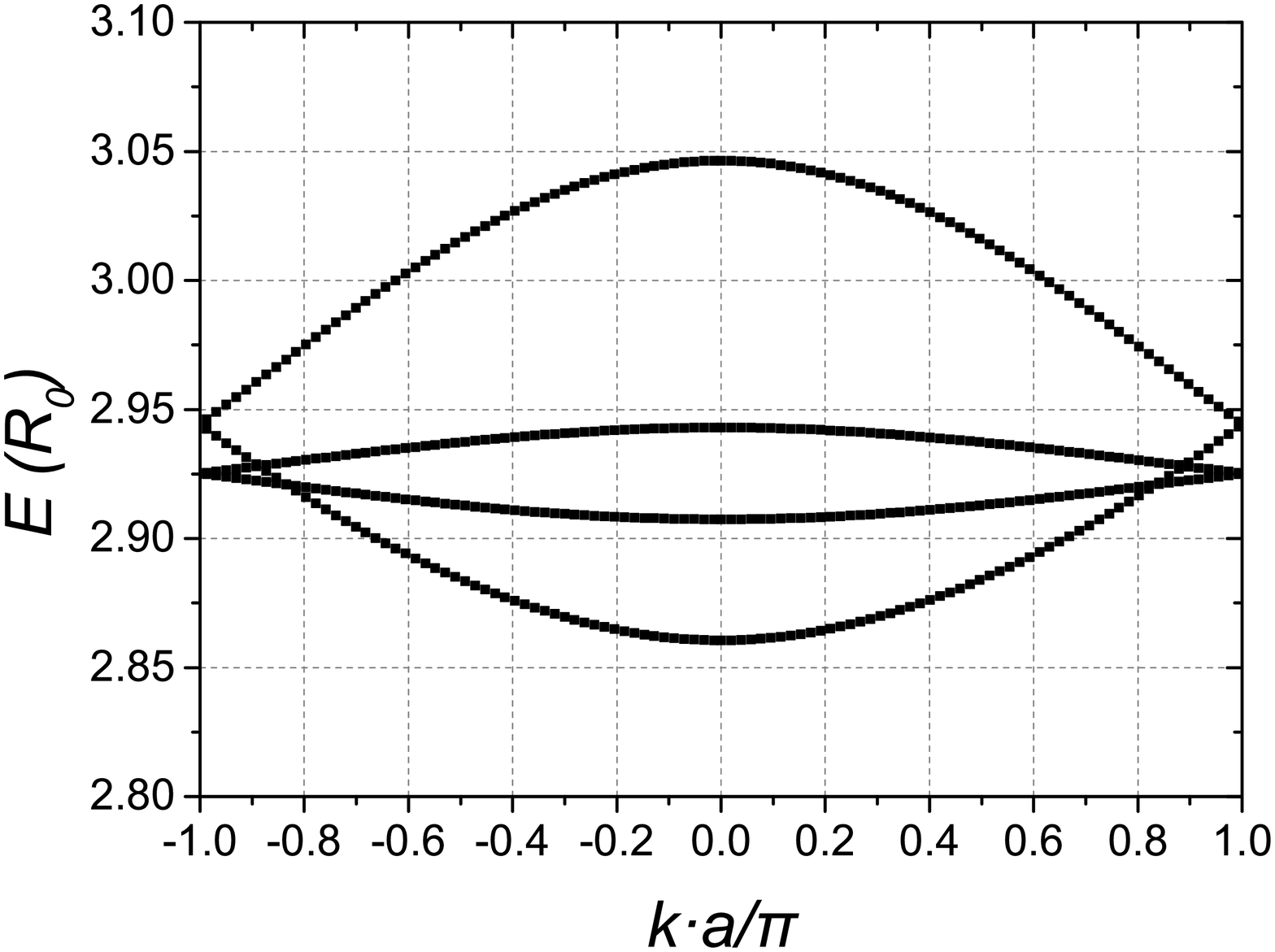}
(g)\includegraphics[scale=0.26]{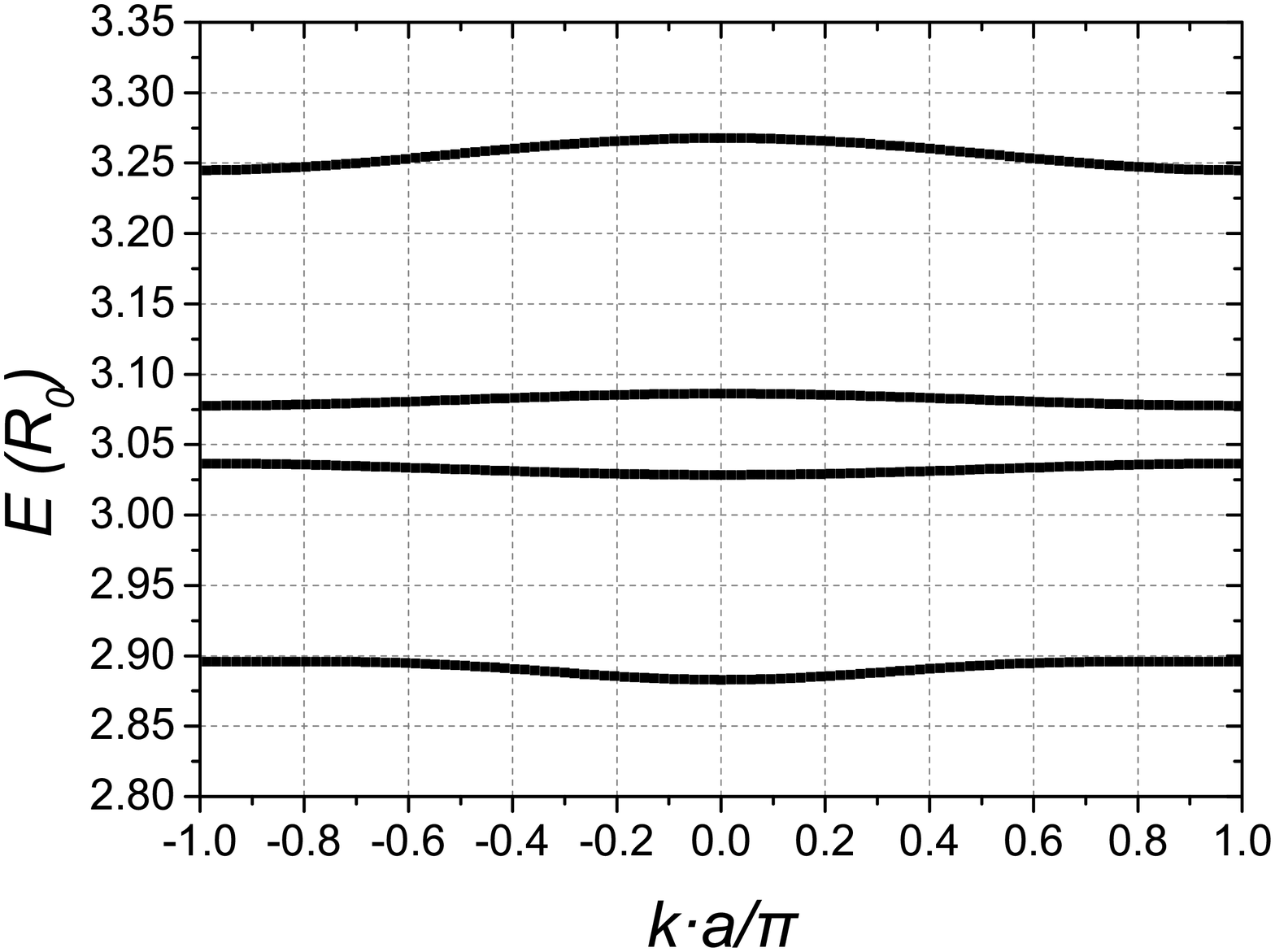}
(h)\includegraphics[scale=0.26]{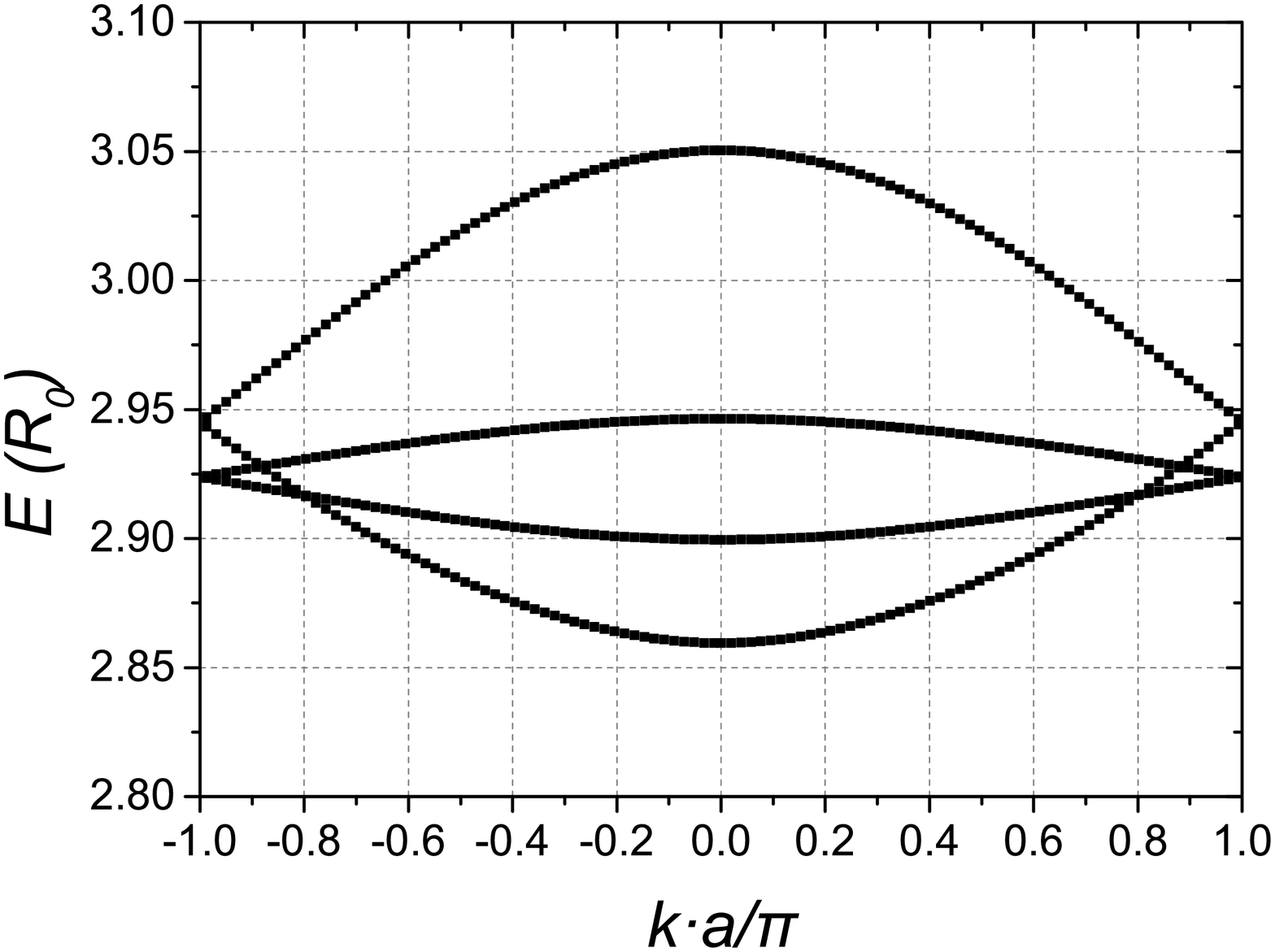}
\caption{The band structure of the lowest 4 energy states for the infinite chain under different arrangements when $d_1+d_2$ is fixed: (a) the spherical case when $d_1=4a_0$ and $d_2=6a_0$, (b) the spherical case when $d_1=d_2=5a_0$, (c) the cubic case when $d_1=2a_0$ and $d_2=4a_0$ in the [001] direction, (d) the cubic case when $d_1=d_2=3a_0$ in the [001] direction, (e) the cubic case when $d_1=2.5a_0$ and $d_2=5a_0$ in the [110] direction, (f) the cubic case when $d_1=d_2=3.75a_0$ in the [110] direction, (g) the cubic case when $d_1=2.5a_0$ and $d_2=5a_0$ in the [111] direction, (h) the cubic case when $d_1=d_2=3.75a_0$ in the [111] direction. For the spherical model calculations, the state with $m_F=3/2$ ($m_F=1/2$) is open (solid).}\label{f-9}
\end{figure*}

For the infinite chain, exchanging the value of $d_1$ and $d_2$ makes no difference on the system. So we only need to consider the short-long arrangement ($d_1\le d_2$) when $d_1+d_2$ is held constant. The band structures under different arrangements for the two spherical cases (long-range and shrort-range) and for the the cubic case in different directions when $d_1+d_2$ is fixed are shown in Figure \ref{f-8} in Appendix \ref{bandstructure}. We also show the detail of the lowest 4 energy bands (those at the top of the graphs in Figure \ref{f-8}) in Figure \ref{f-9}. There are gaps between the bands of states when $d_1\neq d_2$, but these gaps close when $d_1=d_2$, where the periodicity of the model halves and the size of the Brillouin zone doubles.  The calculation could not be done under the condition $d_1+d_2=6a_0$ for the cubic model in the [110] and [111] directions, because the relevant overlap matrix is not positive definite; we use the condition $d_1+d_2=7.5a_0$ instead. 

We now investigate the topological properties of the band structure and their connection to the properties of the finite chains. We calculate the Zak phase $Z$ as described in \S\ref{zakphase}; the results are shown in Table \ref{t-2}. All the short-long arrangement calculations are done under the same conditions as the band structures described above, while the long-short arrangement calculations are done by exchanging the values of $d_1$ and $d_2$. For the cubic case, `first' means that the states correspond to the first curve at the top of the pictures in Figure \ref{f-9}, `second' means that the states correspond to the second curve from the top of the pictures in Figure \ref{f-9}. The results confirm that the states observed to split from the bands in the finite-chain calculations are indeed non-trivial topological states. In general we expect these topological states to arise when the effective chain (after allowing for the split-off of any electrostatically bound states) is terminated by a weak bond; Table \ref{t-2} indeed shows non-trivial Zak phases ($Z=\pi$ mod $2\pi$) for short-long chains; however, the Zak phase for the spherical case with $m_F=\frac{3}{2}$ is abnormal (non-trivial for long-short-chains). 


\begin{table}
\centering
\caption{The Zak phase $Z$ computed under a variety conditions for long-range model in the spherical and cubic cases.}\label{t-2}
\begin{tabular}{|c|c|c|}
\hline
Arrangement&Long-short&Short-long\\
\hline
Spherical case with $m_F=\frac{1}{2}$&$0$&$\pi$\\
\hline
Spherical case with $m_F=\frac{3}{2}$&$-\pi$&$2\pi$\\
\hline
Cubic case in [001] direction (first)&$0$&$\pi$\\
\hline
Cubic case in [001] direction (second)&$0$&$\pi$\\
\hline
Cubic case in [110] direction (first)&$0$&$\pi$\\
\hline
Cubic case in [110] direction (second)&$0$&$\pi$\\
\hline
Cubic case in [111] direction (first)&$0$&$\pi$\\
\hline
Cubic case in [111] direction (second)&$0$&$\pi$\\
\hline
\end{tabular}
\end{table}

Now we show that the existence of `abnormal' values of Zak phase result from the behavior of the effective transition strength between the same single-acceptor level on different nearest-neighbor sites as a function of separation. 

\begin{figure}
\centering
\includegraphics[scale=0.7]{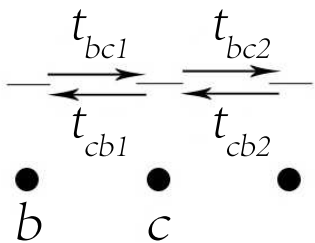}
\caption{Schematic showing the definition of the transition strengths $t_{bc1}$, $t_{cb1}$, $t_{bc2}$, and $t_{cb2}$; here atoms b and c are in the same unit cell.}\label{f-11}
\end{figure}

First, we develop a simple orthogonal 1-level-per-acceptor model in which the only parameters are the transition strengths between different sites, the most important being between nearest neighbors. These are shown in Figure \ref{f-11}: we define $t_{bc1}$, $t_{cb1}$ to be the intra-cell transition strengths, and $t_{bc2}$, $t_{cb2}$ to be the inter-cell transition strengths. Normally, a longer bond length would correspond to a smaller value of the transition strength and a shorter bond to a larger transition energy. But in the `abnormal' case, we find that the dependence is reversed, so the longer bond length has a stronger transition strength for the particular level concerned. This could make the effective arrangement for the system (defined in terms of strong and weak interactions) differ from the geometrical arrangement; hence the system can switch from a `short-long' arrangement to a `long-short' arrangement and \textit{vice versa}. In other words, whether the chain is abnormal or not depends on whether or not its effective arrangement is the same as its geometrical arrangement. 

In the real acceptor chain the states on different sites are in general not orthogonal so we must solve a generalized eigenvalue problem.  This leads us to define an effective transition matrix
\begin{align}
T_{eff}(k)=S^{-\frac{1}{2}}(k)T(k)S^{-\frac{1}{2}}(k).\label{e-a-2}
\end{align}
Under this definition, the eigenvector becomes
\begin{align}
\left|\tilde{u}_k\right\rangle=S^{\frac{1}{2}}(k)\left|u_k\right\rangle\label{e-a-3}
\end{align}
and the Zak phase can be written as
\begin{align}
Z=i\int_{first BZ}dk\left\langle \tilde{u}_k|\partial_k \tilde{u}_k\right\rangle\label{e-a-4}
\end{align}
As shown in a previous paper \cite{Artacho2017QMiaEHS}, the Zak phase remains invariant under the the transformation (\ref{e-a-3}). Therefore we can say the effective transition strength matrices are equivalent to the original transition matrices in the computation of the Zak phase. This argument also shows that Equation (\ref{e-2-4-2}) plays the same role in the generalized eigenvalue problem as Equation (\ref{e-2-4-1}) plays in the case with the orthogonal eigenstates (where $S(k)=1$).

We find that the effective transition strengths can behave differently from the original ones and in particular their dependences on the geometrical arrangement can be opposite. Therefore, once again we need to use an effective arrangement to describe the system, which we define so that the short effective bonds correspond to the strong effective transition strengths, and the long effective bonds to the weak effective transition strengths. With these two new definitions, we find the Zak phase for a particular band is determined by the the effective atomic arrangement; once again, the Zak phase is abnormal when this effective arrangement differs from the actual geometry. An alternative way of phrasing this argument is in terms of the Wannier functions for each band, whose centres of charge are closely related to the Zak phase \cite{Zak1989BPfEBiS} and which are by construction decoupled from the other bands \cite{Marzari1997MN&VDMLGWFfCEBPRB511}.


\section{Conclusion}\label{conclusion}
We have developed an LCAO model to describe the properties of acceptor arrays in tetrahedrally bonded semiconductors, both in the spherical model and the cubic model, within the independent-hole approximation.  We have used it to predict the low-energy states of acceptor dimers and linear acceptor chains in silicon. In particular we have studied the lowest few energy states in the finite chain, arising from linear combinations of the $1S_{3/2}$ (or $1\Gamma_8^+$) acceptor ground states.   For the case of a single hole in the chain we find a complex interplay between the long-range Coulomb interaction and the topological properties of the chain; the electrostatic attraction between the hole and the acceptors in the interior of the chain `splits off' a state localised on the end acceptor, and the transition between topological and non-topological states then takes place in the remainder of the chain.  This has the consequence that a single hole has twofold-degenerate topological bound states derived from the lowest energy band in the `short-long' arrangement (where the chain ends in a short, rather than a long, bond) that merge into the bulk bands in the `long-short' arrangement; these bound states are mainly localised on the \textit{next-to-end} acceptors, and their topological origin can be confirmed by computing the Zak phase in the corresponding infinite chain model. 

In an array with many holes the long-range interactions are likely to be screened out by the motion of other holes.  We approximate this effect by introducing a `short-range model' in which phenomenological screening removes the effect of acceptors beyond the nearest neighbour of each pair.  In this case the electrostatic splitting off of the states localised on the end acceptors disappears, and the topological states of the lowest band (which is derived from the $m_F=\pm\frac{1}{2}$ states in the spherical model) appear for the `long-short' arrangement instead (where the chain ends in a long bond).  The situation in the next-lowest band (derived from the $m_F=\pm\frac{3}{2}$ states in the spherical model) is more complex and we trace this to a non-monotonic dependence of the effective hopping matrix element between these states on the acceptor spacing.  

We note that even with the inclusion of screening, we would not expect our model to be accurate at large spacings (where the Coulomb interactions are expected to dominate over the inter-acceptor tunneling).  For dimerised geometries we would expect the behavior to cross over from a band insulator (at small spacings) to an antiferromagnetic spin model (at large spacings); a similar transition is found in models of donor arrays \cite{nguyen2019}.  The system would, however, remain insulating throughout.  For the equally spaced case ($d_1=d_2$) we would expect a true metal-insulator transition to occur in the real system which, being driven by interactions, is not captured in our model.  Experimental evidence from randomly doped p-type bulk Si suggests this occurs at densities around $4.5\times 10^{18}\,\mathrm{cm}^{-3}$ as shown in the previous paper\cite{Dai1992ECoMSntMIT}, corresponding to spacings around $6\,\mathrm{nm}=2.4\,a_0$; this is within the range of the typical separations ($2a_0$ to $5a_0$) considered in our calculations for the cubic case.  Hence, even when we are working on the insulating side of the transition, our system is relatively close to the phase boundary and we might expect our results to remain qualitatively correct except when $d_1=d_2$ (where we fail to predict the correct insulating behavior).  The cases with $d_1\ne d_2$, showing the topological behavior, should be qualitatively correct.

In conclusion, our results generalise the concept of topological end states to encompass the richness of band-edge degeneracy and spin-orbit coupling expected in acceptor states in silicon.  Our findings point to the complex interplay between topological effects based on the dimerisation, the distance dependence of the interactions, and the long-range electrostatics that is likely to determine the nature and location of the edge states in real acceptor arrays. They also suggest that more complex array geometries beyond simple one-dimensional lines might reveal still richer topological behaviour.

\section{Acknowledgement}

We wish to acknowledge the support of the Engineering and Physical Sciences Research Council and  UK Research and Innovation under the ADDRFSS programme (grant EP/M009564/1). We thank Nguyen Le, Ben Murdin, Neil Curson, and Gabriel Aeppli for helpful and inspiring discussions.

\section*{APPENDIX}

\begin{appendices}

\section{Spherical states}\label{states}
Here we detail the forms of the spherical states with different angular momentum quantum numbers used in the single-acceptor calculations.
\begin{align}
\Phi(S_\frac{3}{2})=f_0(r)\left|L=0,J=\frac{3}{2},F=\frac{3}{2},m_F\right\rangle\notag\\
+g_0(r)\left|L=2,J=\frac{3}{2},F=\frac{3}{2},m_F\right\rangle\notag\\
+h_0(r)\left|L=2,J=\frac{1}{2},F=\frac{3}{2},m_F\right\rangle\label{e-2-1-1-2a}\\
\Phi(S_\frac{1}{2})=f_1(r)\left|L=0,J=\frac{1}{2},F=\frac{1}{2},m_F\right\rangle\notag\\
+g_1(r)\left|L=2,J=\frac{3}{2},F=\frac{1}{2},m_F\right\rangle\label{e-2-1-1-2b}\\
\Phi(P_\frac{1}{2})=f_2(r)\left|L=1,J=\frac{3}{2},F=\frac{1}{2},m_F\right\rangle\notag\\
+g_2(r)\left|L=1,J=\frac{1}{2},F=\frac{1}{2},m_F\right\rangle\label{e-2-1-1-2c}\\
\Phi(P_\frac{3}{2})=f_3(r)\left|L=1,J=\frac{3}{2},F=\frac{3}{2},m_F\right\rangle\notag\\
+g_3(r)\left|L=1,J=\frac{1}{2},F=\frac{3}{2},m_F\right\rangle\notag\\
+h_3(r)\left|L=3,J=\frac{3}{2},F=\frac{3}{2},m_F\right\rangle\label{e-2-1-1-2d}\\
\Phi(P_\frac{5}{2})=f_4(r)\left|L=1,J=\frac{3}{2},F=\frac{5}{2},m_F\right\rangle\notag\\
+g_4(r)\left|L=3,J=\frac{3}{2},F=\frac{5}{2},m_F\right\rangle\notag\\
+h_4(r)\left|L=3,J=\frac{1}{2},F=\frac{5}{2},m_F\right\rangle\label{e-2-1-1-2e}
\end{align}
Here $f_i(r)$, $g_i(r)$ and $h_i(r)$ are the radial parts and $\left|L,J,F,m_F\right\rangle$ is the angular part, which can be expanded in terms of Gaussian functions \cite{Baldereschi1973SMoSASiS}. The further spherical states we need to form the cubic states can be obtained in the same way.

\section{Transition elements}\label{transitionelement}
Here we show the transition elements used in the finite chain calculation. The subscript indicates the acceptor on which the state involved in the transition is centred.
\begin{align}
&t_{aa}=E_{a}S_{aa}+V_{b}\label{e-a-1a}\\
t_{ab}=\frac{1}{2}E_{a}&S_{ab}+\frac{1}{2}E_{b}S_{ab}+\frac{1}{2}V_{a}+\frac{1}{2}V_{b}+V_{c}\label{e-a-1b}\\
t_{ac}=\frac{1}{2}E_{a}S_{ac}&+\frac{1}{2}E_{c}S_{ac}+\frac{1}{2}V_{a}+V_{b}+\frac{1}{2}V_{c}+V_{d}\label{e-a-1c}\\
t_{bb}&=E_{b}S_{bb}+V_{a}+V_{c}\label{e-a-1d}\\
t_{bc}=\frac{1}{2}E_{b}S_{bc}&+\frac{1}{2}E_{c}S_{bc}+V_{a}+\frac{1}{2}V_{b}+\frac{1}{2}V_{c}+V_{d}\label{e-a-1e}\\
t_{bd}=\frac{1}{2}E_{b}S_{bd}+&\frac{1}{2}E_{d}S_{bd}+V_{a}+\frac{1}{2}V_{b}+V_{c}+\frac{1}{2}V_{d}+V_{e}\label{e-a-1f}
\end{align}
Here $E_{i}$ is the single-acceptor energy of the state on atom $i$, $S_{ij}$ is the overlap matrix between the states on atom $i$ and atom $j$, $V_{i}$ is the potential matrix of atom $i$. Here $t_{aa}$, $t_{ab}$ and $t_{ac}$ are for the acceptor at the end of the chain, $t_{bb}$, $t_{bc}$ and $t_{bd}$ are for the acceptor in the middle of the chain. $t_{bb}$, $t_{bc}$ and $t_{bd}$ are also used in the infinite chain calculation.

\section{$\mathrm{H}^{+}_{2}$ case}\label{h2+}
\begin{figure}
\centering
\includegraphics[scale=0.28]{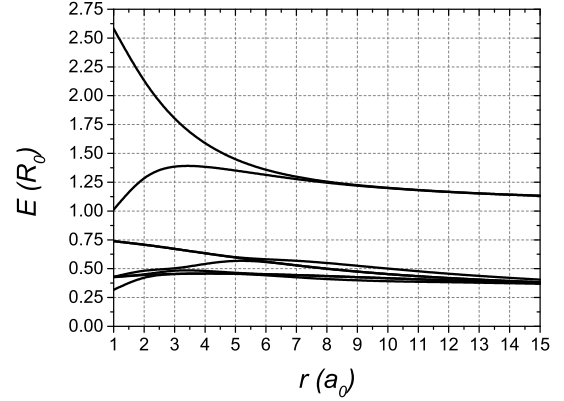}
\caption{The behavior of the eigenenergies as a function of separation $r$ for the $\mathrm{H}^{+}_{2}$ molecular ion.}\label{f-10}
\end{figure}

The $\mathrm{H}^{+}_{2}$ molecular ion is an analogue of the one-hole, two-acceptor problem but without any spin-orbit coupling or heavy-hole light-hole splitting ($\mu=\Delta=\delta=0$, and $\gamma=1$).  We show the behavior of the eigenenergies as a function of separation $r$ for the $H^{+}_{2}$ case, computed in the same way as the corresponding spherical acceptor problem, in Figure \ref{f-10}. The results show the expected splitting of the atomic 1s states into bonding and antibonding orbitals, and also the splitting of the 2s and 2p levels as the atomic spacing is reduced, but without the spin-orbit splittings present in the acceptor analogues.  

\section{Eigenvalues of the overlap matrix}\label{overlapeigenvalues}
\begin{figure}
\centering
\includegraphics[scale=0.28]{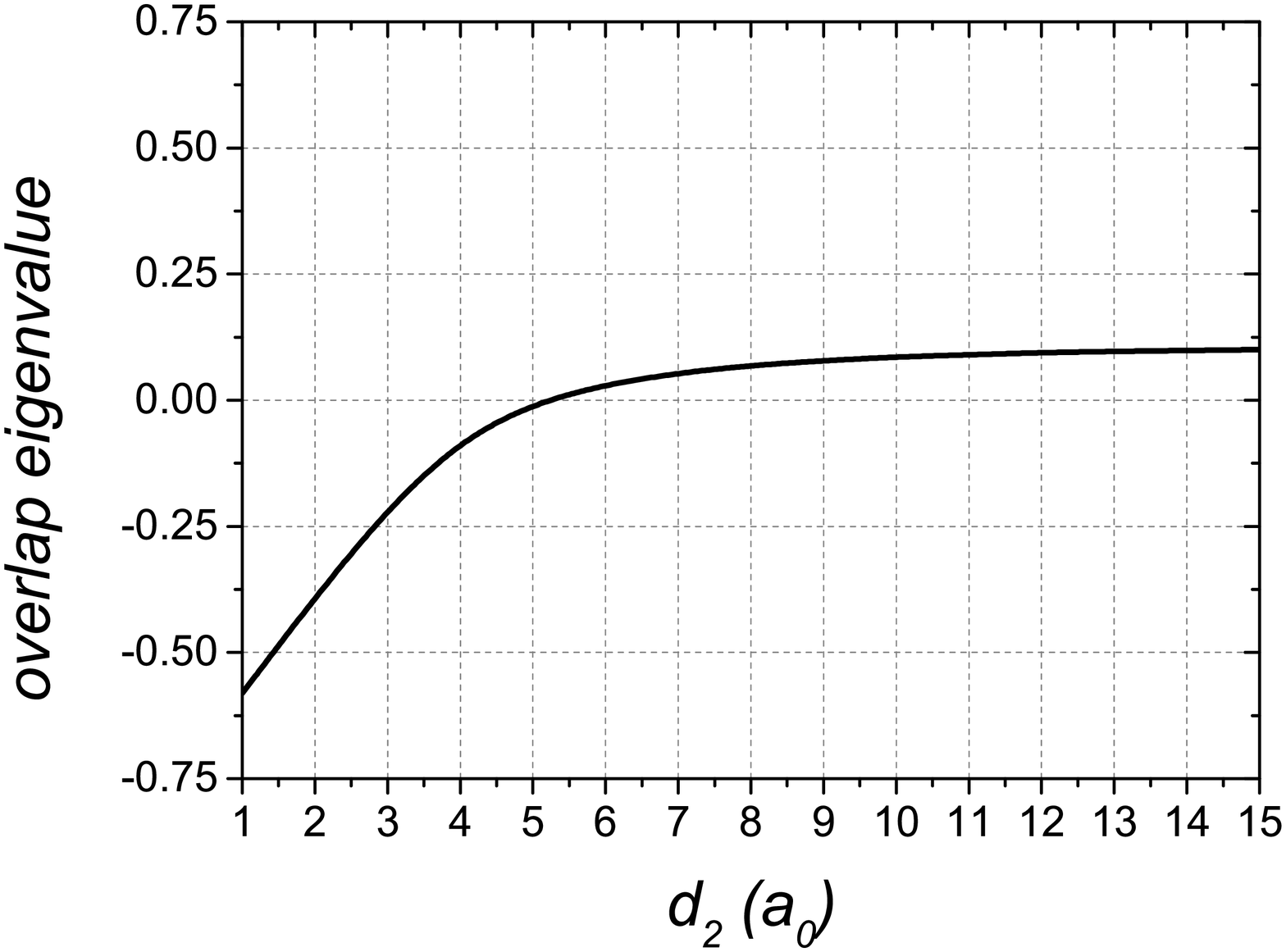}
\caption{The smallest eigenvalue of the overlap matrix, truncated at next-nearest-neighbors, for a 10-acceptor finite chain in the spherical model with $m_F=\frac{1}{2}$ as a function of separation $d_2$ when $d_1=4a_0$.  Note the appearance of unphysical negative eigenvalues when $d_2\le 6a_0$.}\label{f-5}
\end{figure}

Here we show the behavior of the smallest eigenvalue of the overlap matrix for a 10-acceptor finite chain under the spherical model when $m_F=\frac{1}{2}$ with different separation $d_2$ when $d_1=4a_0$ in Figure \ref{f-5}. It can be seen that the negative eigenvalue disappear for $d_2\ge6a_0$. Similar thresholds can be found for other values of $d_1$.

\section{Band structures for the infinite chain}\label{bandstructure}
\begin{figure*}
\centering
(a)\includegraphics[scale=0.26]{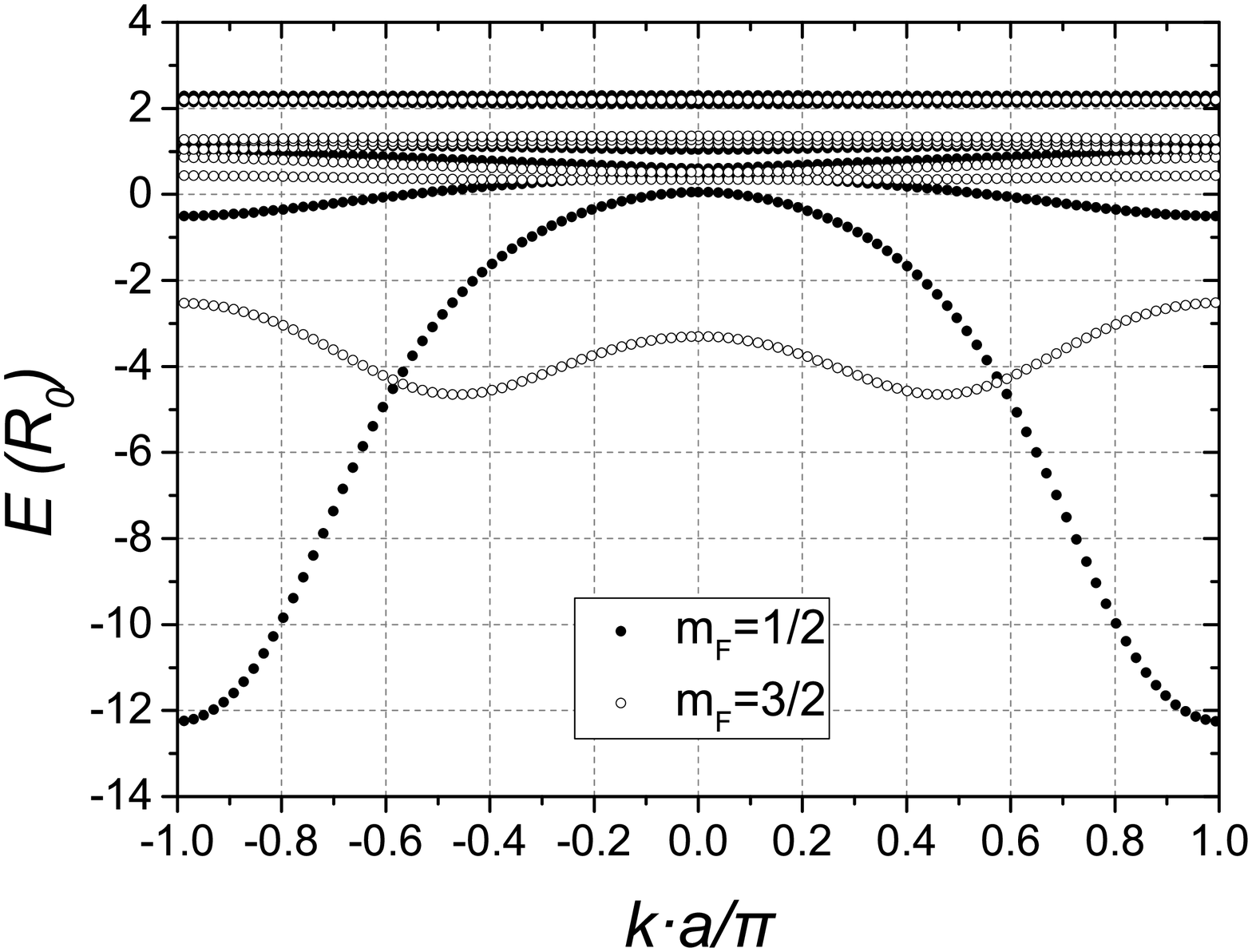}
(b)\includegraphics[scale=0.26]{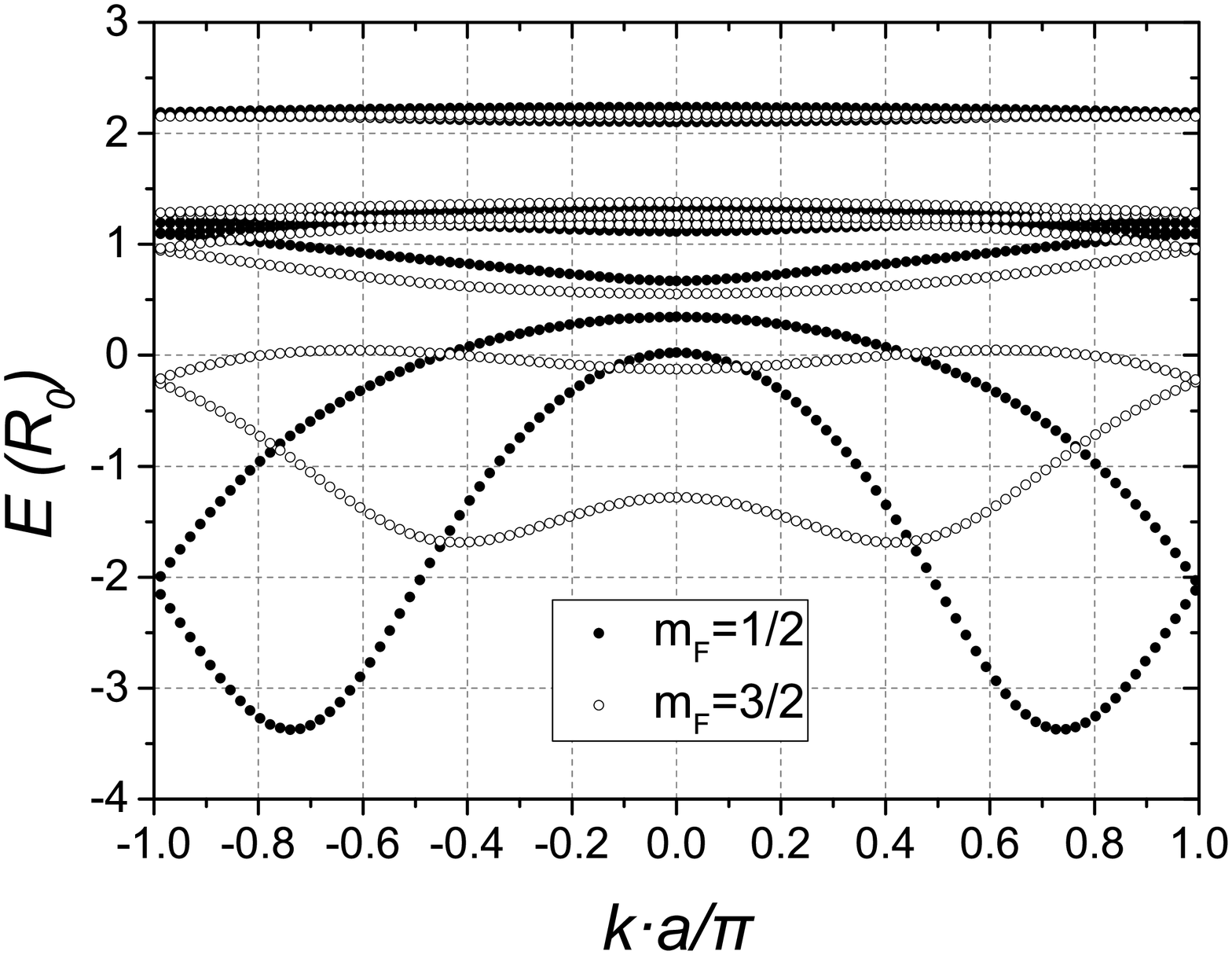}
(c)\includegraphics[scale=0.26]{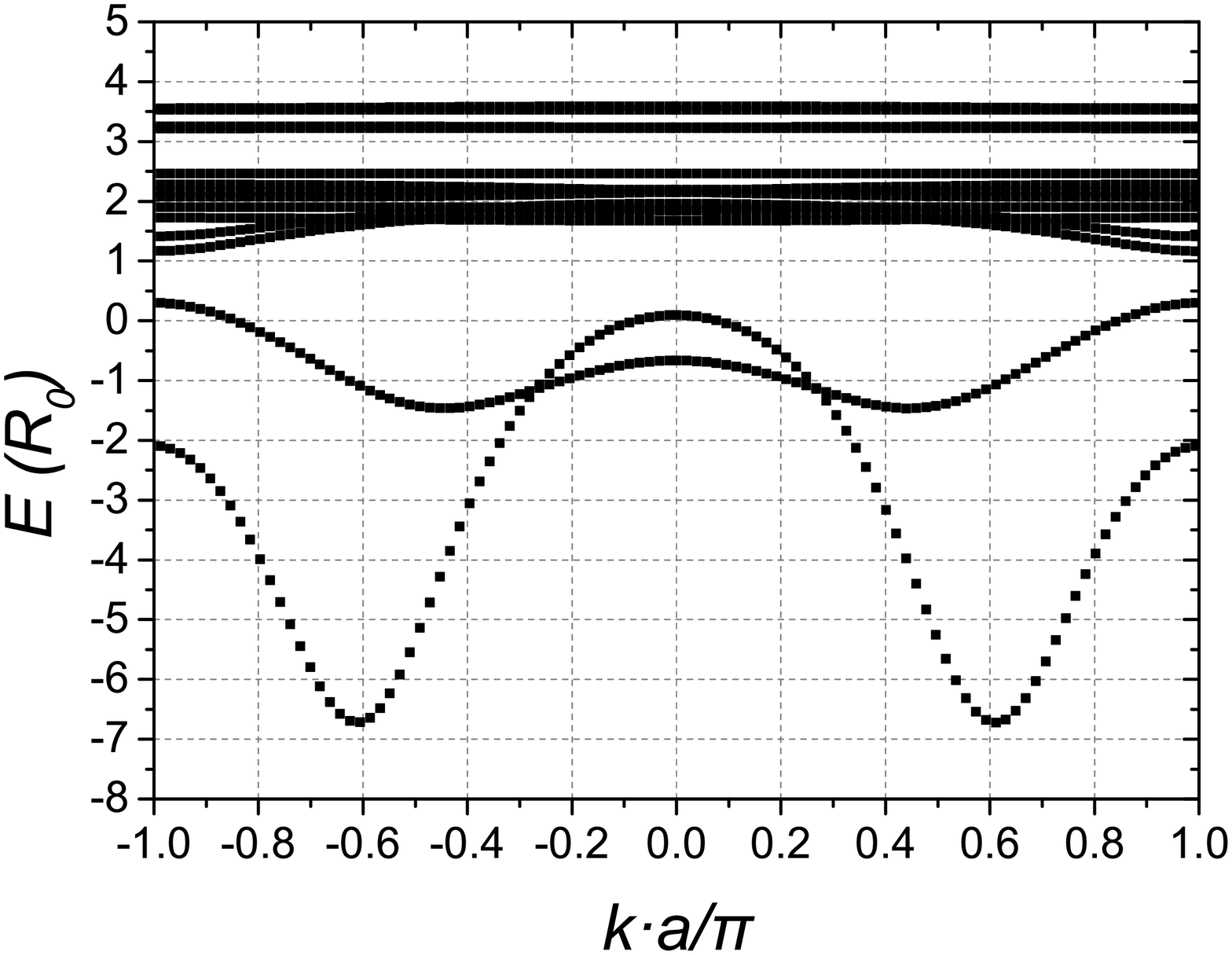}
(d)\includegraphics[scale=0.26]{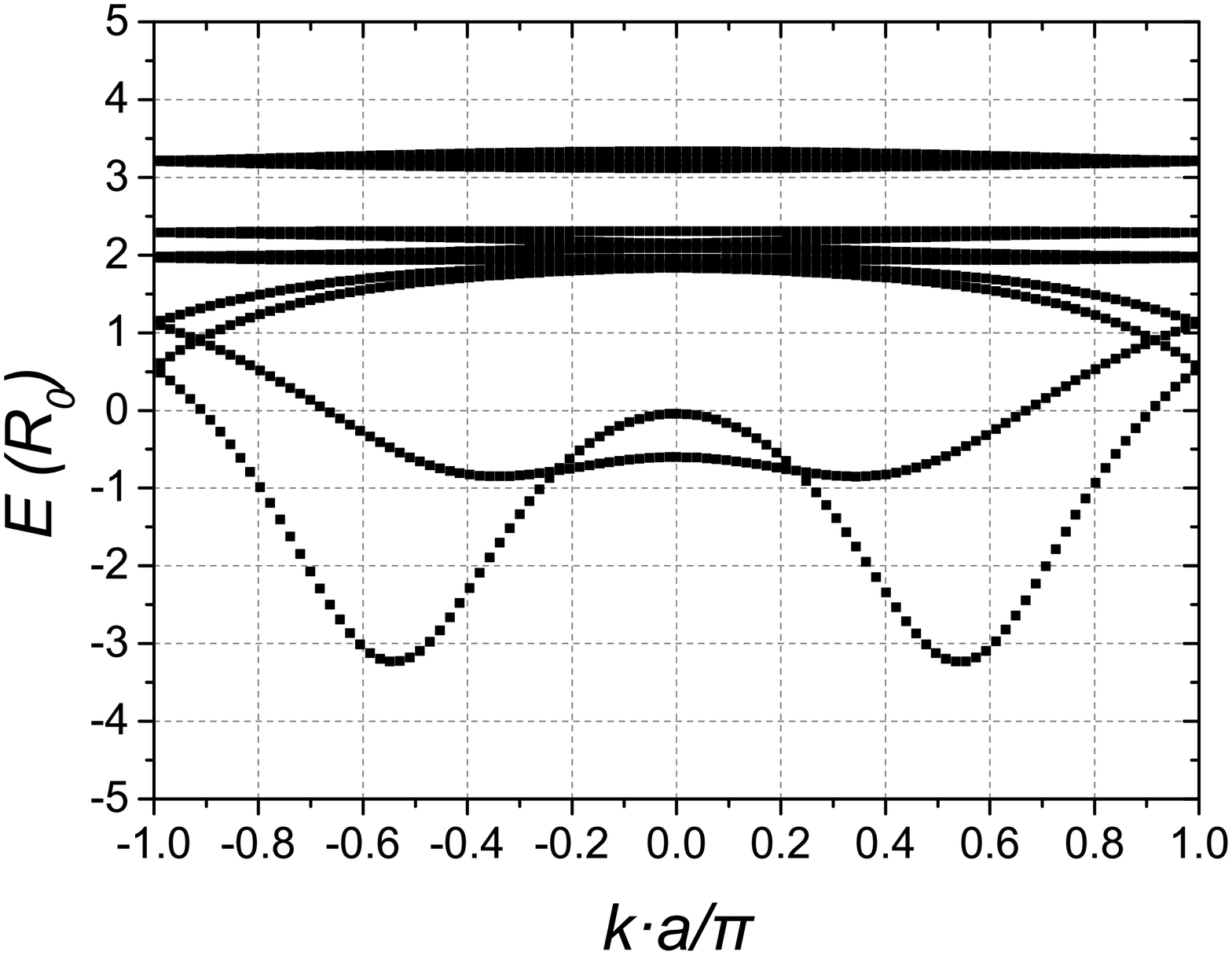}
(e)\includegraphics[scale=0.26]{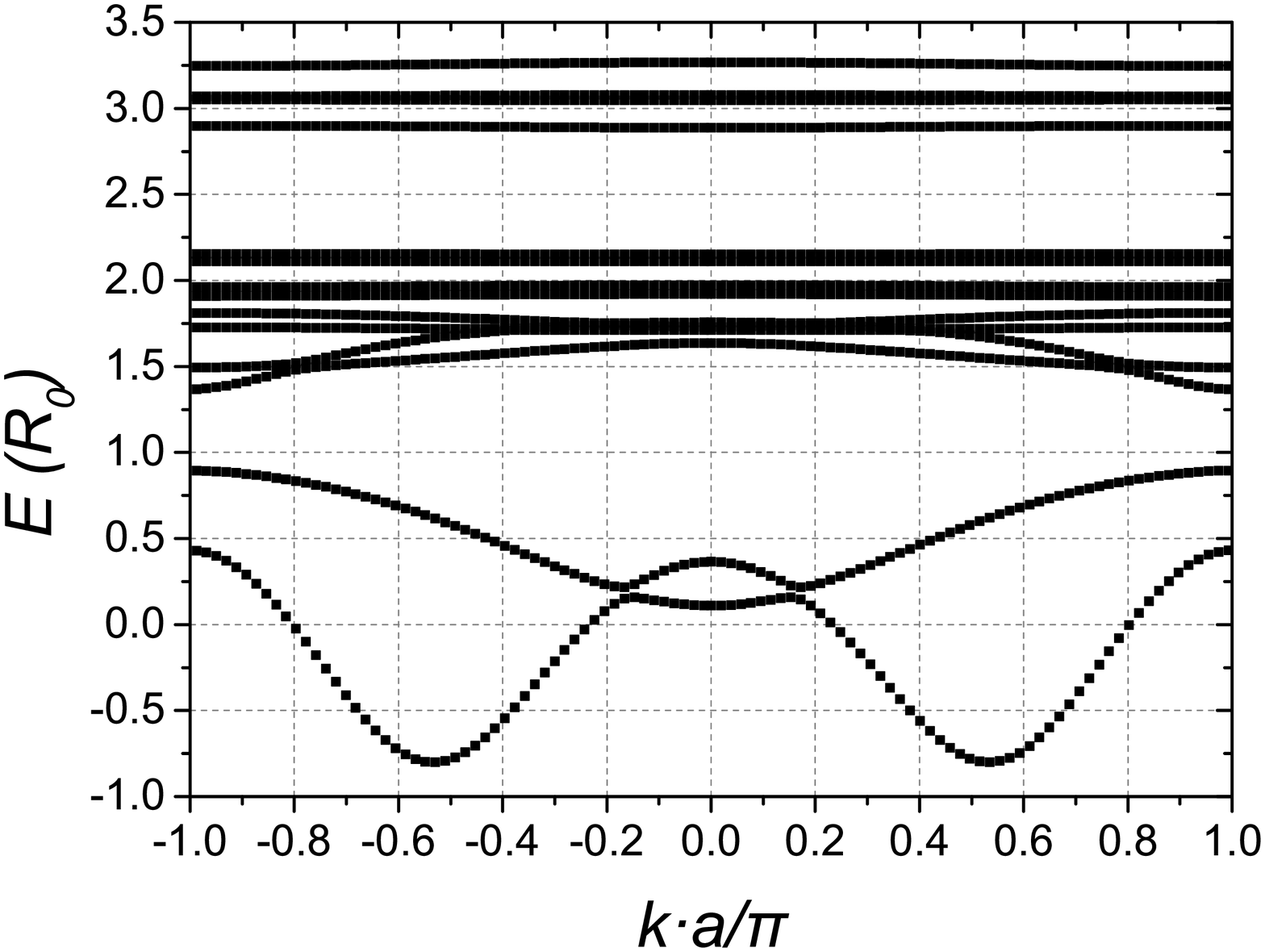}
(f)\includegraphics[scale=0.26]{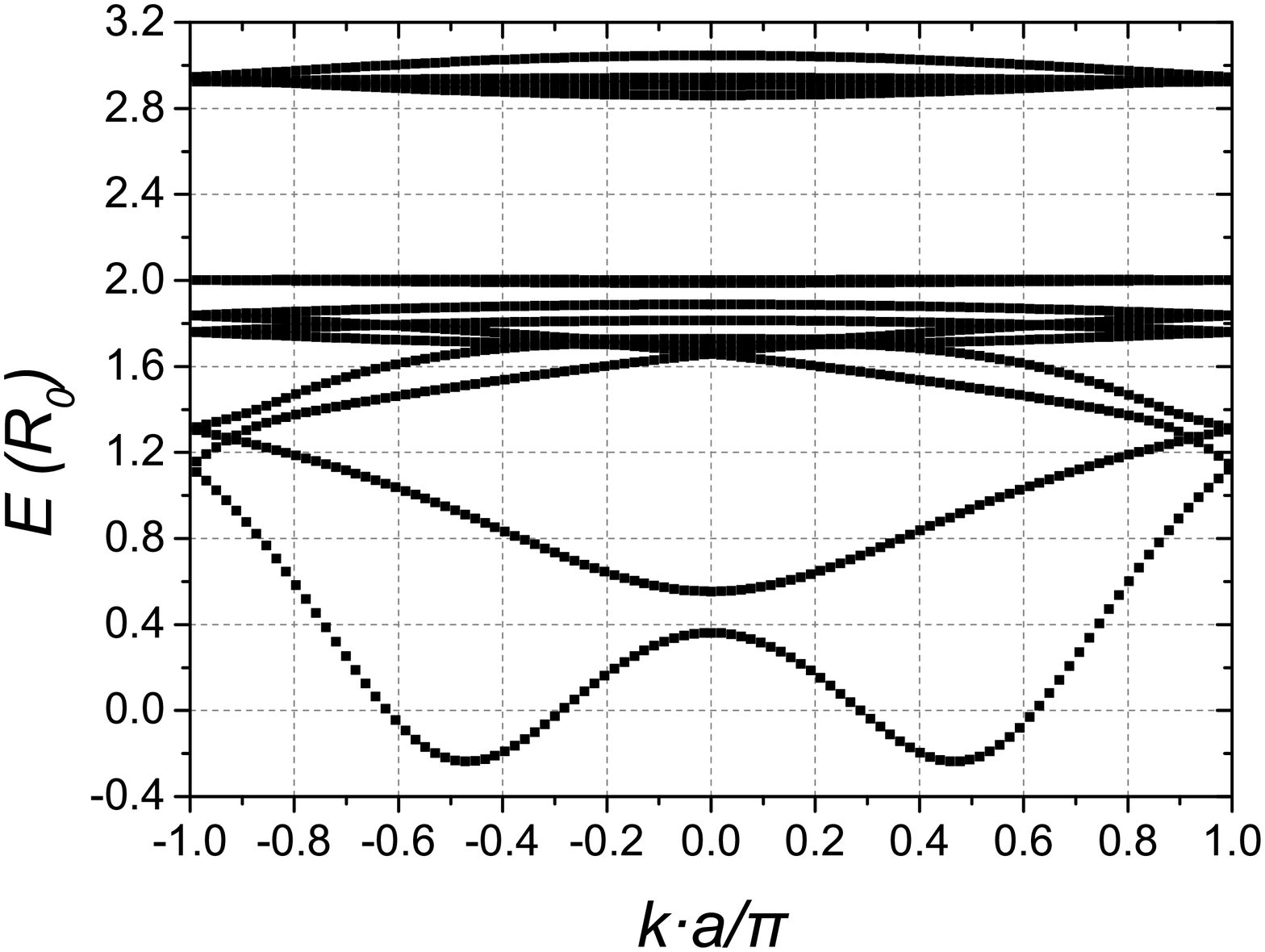}
(g)\includegraphics[scale=0.26]{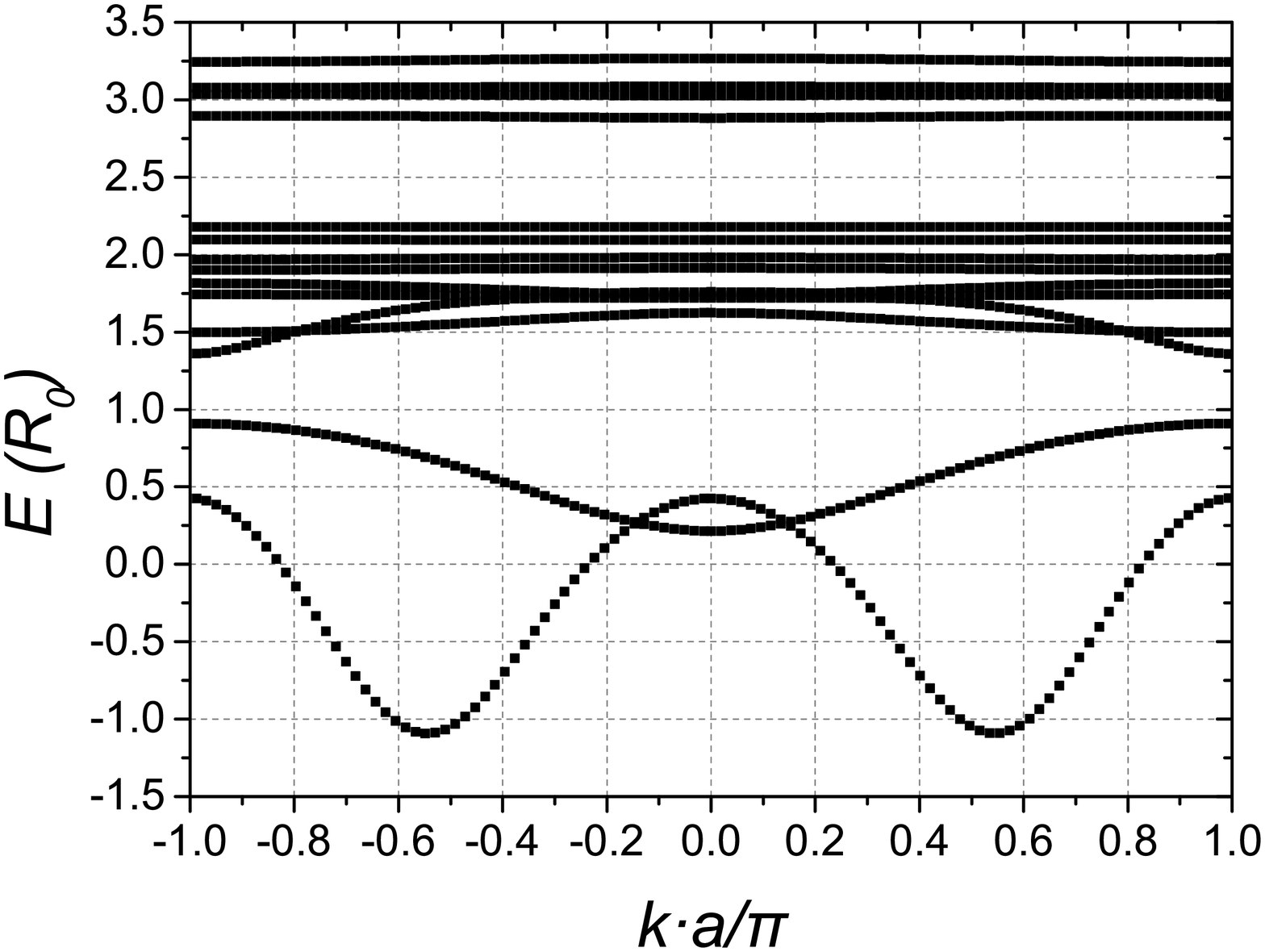}
(h)\includegraphics[scale=0.26]{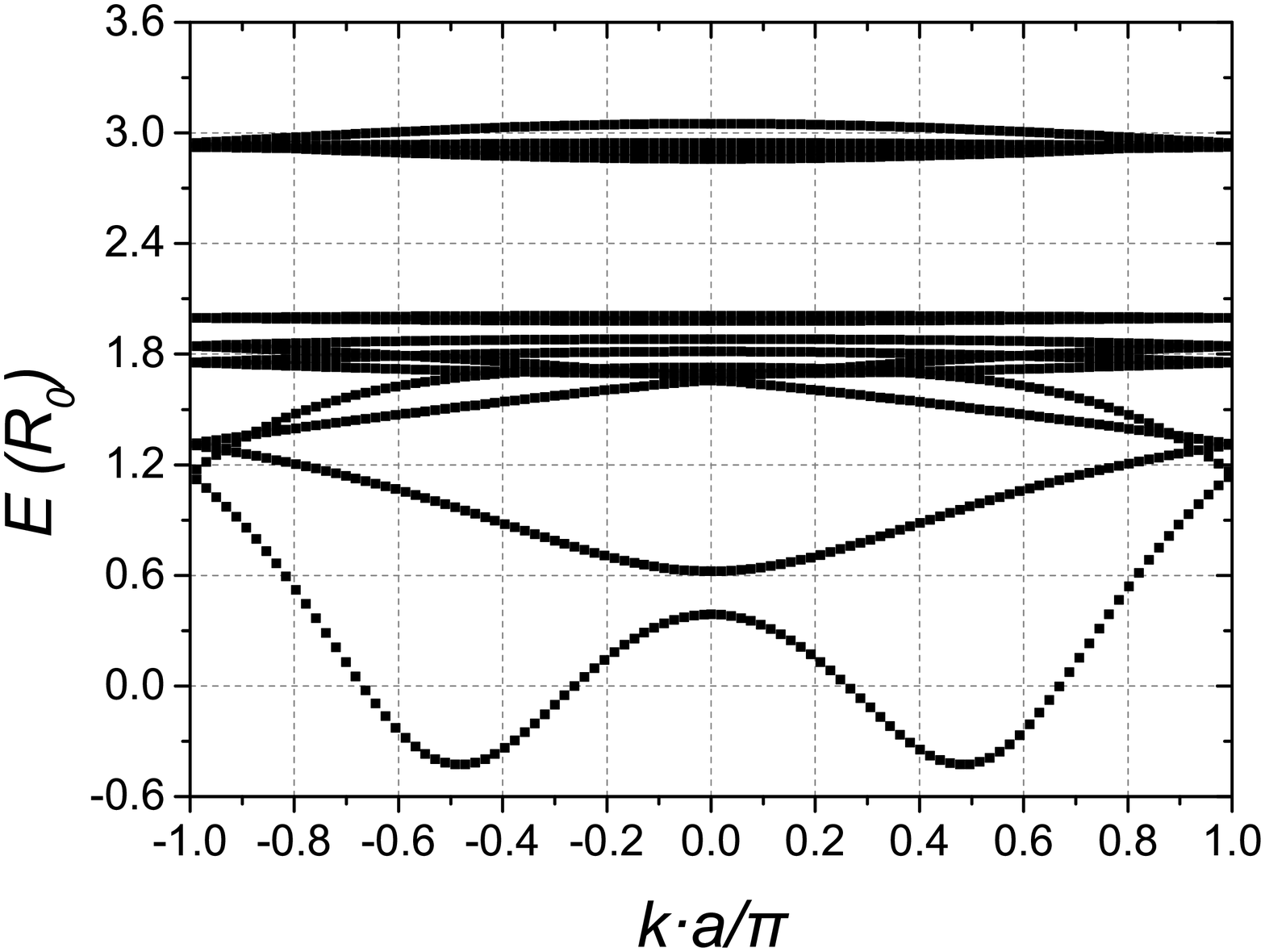}
\caption{The band structure for the infinite chain under different arrangements when $d_1+d_2$ is fixed. (a) the spherical case when $d_1=4a_0$ and $d_2=6a_0$, (b) the spherical case when $d_1=d_2=5a_0$, (c) the cubic case when $d_1=2a_0$ and $d_2=4a_0$ in the [001] direction, (d) the cubic case when $d_1=d_2=3a_0$ in the [001] direction, (e) the cubic case when $d_1=2.5a_0$ and $d_2=5a_0$ in the [110] direction, (f) the cubic case when $d_1=d_2=3.75a_0$ in the [110] direction, (g) the cubic case when $d_1=2.5a_0$ and $d_2=5a_0$ in the [111] direction, (h) the cubic case when $d_1=d_2=3.75a_0$ in the [111] direction. For the spherical model calculations, the state with $m_F=3/2$ ($m_F=1/2$) is open (solid).}\label{f-8}
\end{figure*}

Here we show the band structure under different arrangements when $d_1+d_2$ is fixed in Figure \ref{f-8}.  Note that the bands cross in the [111] direction while they anticross in the [110] direction.  This difference can be traced back to the symmetry properties of the different geometries (see Table~\ref{t-3}): different bands can have the different symmetries at the same $k$ in the [111] direction, and therefore the bands can cross.  In the [110] direction, however, all bands have the same symmetry and so they anticross.

\end{appendices}

\bibliography{ref}

\end{document}